\documentclass{aa}  
\usepackage{graphicx}
\usepackage{natbib}

\bibpunct{(}{)}{;}{a}{}{,} % to follow the A&A style
\usepackage[colorlinks=true,linkcolor=blue,citecolor=blue,urlcolor=blue]{hyperref}%

\usepackage{txfonts}
\begin{document} 
   \title{Spectroscopic characterization of
   a remarkable temporally varying, triple-lensed quasar at $z=2.67$}
   % {\em Gaia}-assisted

   \author{Charlie~Lind-Thomsen
          \inst{1}\fnmsep\inst{2}          
          \and 
          Kasper~E.~Heintz
          \inst{1}\fnmsep\inst{2}
          \and Albert~Sneppen
          \inst{1}\fnmsep\inst{2}
          \and Kostas~Valeckas
    \inst{1}\fnmsep\inst{2}\fnmsep\inst{4}
          \and Stefan~Geier\inst{3}
          \and Jens-Kristian~Krogager\inst{6}\fnmsep\inst{5}
          \and Johan~Richard\inst{5}
          \and Johan~P.~U.~Fynbo 
          \inst{1}\fnmsep\inst{2}
}

   \institute{Cosmic Dawn Center (DAWN),
         \and
             Niels Bohr Institute, University of Copenhagen, Jagtvej 128, DK-2200, Copenhagen N, Denmark
        \and
            Institut für Physik und Astronomie, Universität Potsdam, Haus 28, Karl-Liebknecht-Str. 24/25, 14476 Potsdam-Golm, Germany
        \and
            GRANTECAN, Cuesta de San Jos\'{e} s/n, 38712 Bre{\~n}a Baja, La Palma, Spain Instituto de Astrof\'{\i}sica de Canarias, V\'{\i}a   L\'actea, 38205 La Laguna, Tenerife, Spain
        \and 
            Centre de Recherche Astrophysique de Lyon, Université de Lyon 1, UMR5574, 69230 Saint-Genis-Laval, France
        \and 
            French-Chilean Laboratory for Astronomy, IRL 3386, CNRS and U. de Chile, Casilla 36-D, Santiago, Chile
}

   %\date{Received DATE, 2025; accepted March 16, 1997}

% \abstract{}{}{}{}{} 
% 5 {} token are mandatory
 
  \abstract
   {Gravitationally lensed quasars are viable cosmic tools for constraining a diversity of fundamental astrophysical phenomena; They enable identification of faint, low-mass supermassive black holes, provide unique constraints on the intervening intergalactic or interstellar medium in their sightlines, and can be used to determine key cosmological quantities such as the Hubble constant, $H_0$. However, they are rare phenomena, and it has proven difficult to define efficient, unbiased selection methods.}
   %work as a constraint on the formation of early SMBHs and a trace of the gas and dust content of the ISM. However, they are rare and faint phenomena, and it has proven hard to find effective selection criteria to find them.}
    {In this study, we report the independent spectroscopic identification of a remarkable triple-lensed quasar system at $z=2.67$, identified based on its astrometric measurements from the {\em Gaia} mission, previously identified using neural network on Pan-STARRS. Furthermore, a larger spectroscopic follow-up survey of {\em Gaia}-detected candidate lensed quasars.}
   %{In this study, we do follow up spectroscopy on a catalogue of astrometrically selected quasars to evaluate a catalogue of astrometrically selected lensed quasar candidates. Furthermore, we do a detailed study of an interesting lensed quasar found in this process.}
   {We characterize in detail the three mirror images of the quasar and their spatial and temporal spectroscopic coverage, with focus on the emission-line properties which shows variation across sigthlines and temporal evolution over the $\sim 11$months  spectroscopic campaign.}
   %{The lens candidates are either accepted or declined as lenses, or they keep their lens candidate status based on their spectrum. One interesting find, the lensed quasar DR3Gaia2107-1611, is studied in detail. A lens model is constructed from a combination of spectrum and imaging.}
   {We construct a lens model of the foreground source from a combination of the multiple spectra and deep optical imaging, providing a robust halo mass of $M_{\rm h} = (2.78 \pm 0.05)\times 10^{10}M_\odot$. Based on the lens model, the time delay between each sightline is translated into an intrinsic quasar time, allowing us to construct a quasar timeseries over $\sim18$months with monthly cadence. Over months timescales the broad emission lines vary in both velocity offset and equivalent width (EW) as well as an overall increase in ionization.}
   {This exemplary triple-lensed quasars demonstrates the viability of identifying such rare lens configurations based purely on the astrometric measurements from the {\em Gaia} mission, which we here provide optimized selection criteria for, for future studies. }
   %This observation of a remarkable quasar lens as a benchmark for defining new, optimized selection strategies with {\em Gaia} and conclude on the viability of selecting these rare systems using astrometry alone. Furthermore, we report on the interesting observation of a triple lensed quasar, displaying both microlensing and temporally varying emission lines. The latter is thought to arise either from an intrinsic change in the quasar itself or from a shift in which region of the quasar dominates along the line of sight.}


   \maketitle
%
%-------------------------------------------------------------------

\section{Introduction}
\label{sec:int}
 Gravitationally lensed quasars are of great observational and theoretical interest; high redshift lensed quasars put the most stringent constraints on the formation of supermassive black holes \citep[SMBHs; see e.g.][]{2012_Volonteri}, allow detailed studies of the chemical abundances in the interstellar medium (ISM) of foreground galaxies \citep[e.g.,][]{2018_Krogager}, and provide independent measures of the Hubble constant, $H_0$, based on the time delay between the different lines of sight and the mass distribution of the lens \citep[e.g.][]{2004_H0_from_lens, 2015_H0_from_lens, 2020_H0_from_lens}. However, gravitationally lensed quasars are rare, faint, and hard to distinguish from other astrophysical objects due to the diversity in emission profiles depending on what angle the quasar is observed at (i.e. whether a jet is in the line of sight), diversity in dust-reddening and redshift dependent color-selections, and the potential small separations of the lensed images.
 Thus, we know very few of these objects, with the most complete catalog containing just $\sim 300$ verified lensed quasars \citep[see e.g. the compilation by the][]{cambridge_lensed_quasars}. It is therefore crucial to develop new selection methods to identify these rare systems in a complete and unbiased way. The objective of this work is to evaluate a specific case of astrometric selection criteria to select lensed quasar candidates and to study any interesting specific cases.

% Quasar selection
The varying spectral properties of quasar due to angle variations, combined with the diversity in redshift and dust extinction, makes it difficult to define complete and unbiased criteria for quasar candidates, as various criteria have different color-biases and intrinsic emission mechanisms for the selected quasars. Especially high-redshift or dust-reddened quasars are not included in studies using color-selection as a criterion for finding quasars \citep{2015_Krawczyk, 2018_Krogager, Heintz_2018, 2019_Krogager}, in particular the typical UV excess selection adopted for the Sloan Digital Sky Survey \citep[SDSS;][]{2004_Richards}.

\begin{figure*}
    \centering
    \includegraphics[width=1\linewidth]{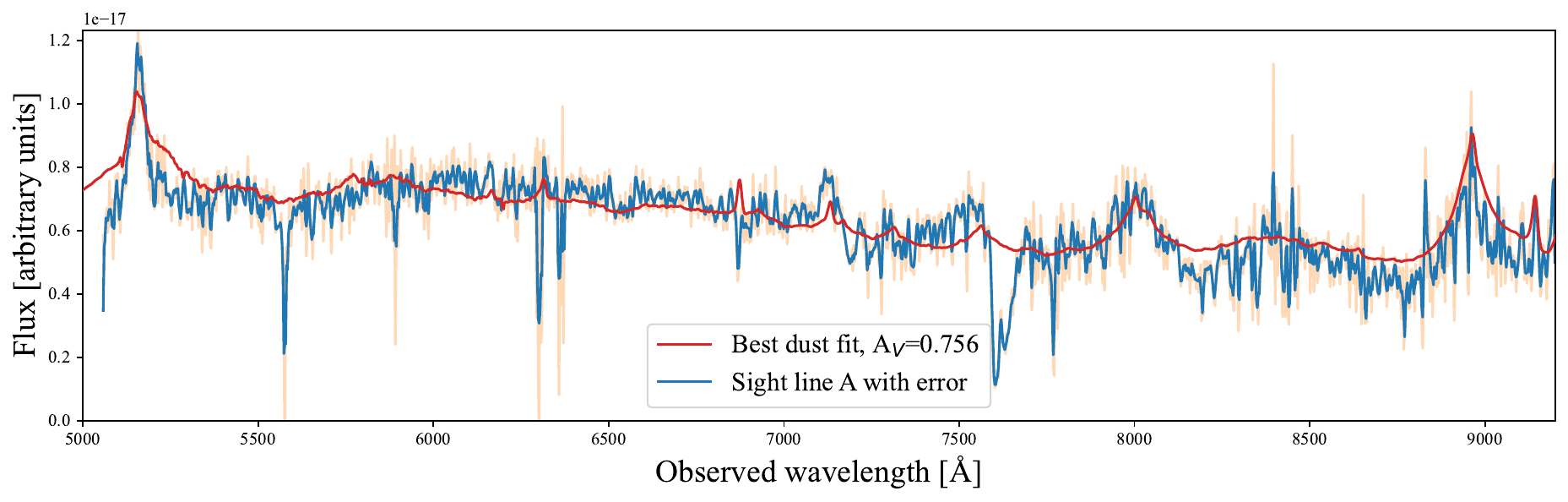}
    \caption{Example spectrum of the quasar DR3Gaia0205-3233 at $z=0.843\pm0.003$ (blue), fit with the dust extinction parameter, $A_V$, red (see Eq. \ref{eq:dust_model}). The redshift is determined from the identified emission lines in the spectrum. }
    %Although only the continuum is part of the fit, several emission lines are reasonable close to template predictions.}
    \label{fig:dust_fit}
\end{figure*}

Identifying quasars based purely on astrometric criteria was first proposed by \citet{1986_Koo}. This method relies only on the zero proper motions of the quasars on their sky due to their large cosmological distances \citep[see also][in context of the {\em Gaia} mission]{2015_Heintz}. Thus, it is complete down to the limiting flux of the instrument and no color criteria are needed. This would otherwise infer a bias in the type of quasars found, as the emission depends highly on the angle from which the quasar is viewed along the line of sight.Motivated by this, \citet{Heintz_2018, Heintz_2020} examined the complete population of objects found in the {\em Gaia} data release 2 \citep[DR2;][]{2018_GaiaDR2} within $1^\circ$ of the Galactic North Pole, based solely on the lack of proper motions of the brightest targets in the field with $G < 20 \, \mathrm{mag}$.
%with with proper motion signal-to-noise ratio of $S_\mu/N<2\sigma$ and brightness in the Gaia G ($\sim$330 nm - 1050 nm) band, $G < 20 \mathrm{mag}$, the latter is simply to remove noise, as this is the limiting magnitude of Gaia. 
The complete spectroscopy coverage of all the quasar candidates with zero proper motions enabled the identification of quasars with an efficiency of $\approx60\%$. Since this method is unbiased in terms of the intrinsic spectrum of the object, this selection provides a complete census of the quasar population within the limiting magnitude of {\em Gaia}. The efficiency of other quasar selection techniques using optical, near- or mid-infrared photometry \citep{2015_Secrest, 2012_Stern, 2012_Mateos, 2008_Maddox, 2000_Warren} is found to be between $85\%-90\%$. Thus, using purely astrometric selection criteria has proved to provide the most complete and unbiased sample of quasars, though at the cost of relatively low purity, in particular at low Galactic latitudes where high stellar contamination is found.
The all-sky coverage of {\em Gaia} with precise astrometric data \citep{Gaia_2023} further enables an efficient selection of quasars across the entire sky. 

%\begin{figure}
%     \centering
%     \includegraphics[width=1\linewidth]{Pictures/spec_box_with_galaxy.png}
%     \caption{Example of a small portion of a reduced DR3Gaia2107-1611 2D spectra recorded by GTC OSIRIS, with the lensing galaxy spectrum visible in the middle. The figure illustrates the close proximity of the 2D traces, constraining the extraction to a fixed region (marked with red) in order to avoid contaminating the 1D extracted spectra with flux from neighbouring objects.}
%     \label{fig:spec_overlap_example}
% \end{figure}

% Note, that the Gaia data catalogue is superior in regard to determining populations through astrometric criteria, as it contains the largest all-sky collection of astrophysical objects with the most precise astrometric data \cite{Gaia_2023}, compared to other all-sky surveys e.g. SDSS, Pan-STARRS, etc. For the study of lenses, a high resolution is essential, due to the small separations between the sight lines. In this regard, Gaia is also superior, with a resolution of $\sim0.6\ \mathrm{arcsec}$ \cite{Gaia_2023}, compared to the resolution of the SDSS, $\sim1.2\ \mathrm{arcsec}$ \cite{2002_SDSS}, and Pan-STARRS, $\sim1.2\ \mathrm{arcsec}$ \cite{chambers2019panstarrs1surveys}, $\sim2.5\ \mathrm{arcsec}$ for 2MASS \cite{2006_2MASS} and $\sim6\ \mathrm{arcsec}$ for AllWise \cite{2014_AllWise}. Thus, Gaia is also superior in separating sources, critical in the selection of lenses.

\begin{figure}[t]
    \centering
    \includegraphics[width=1\linewidth]{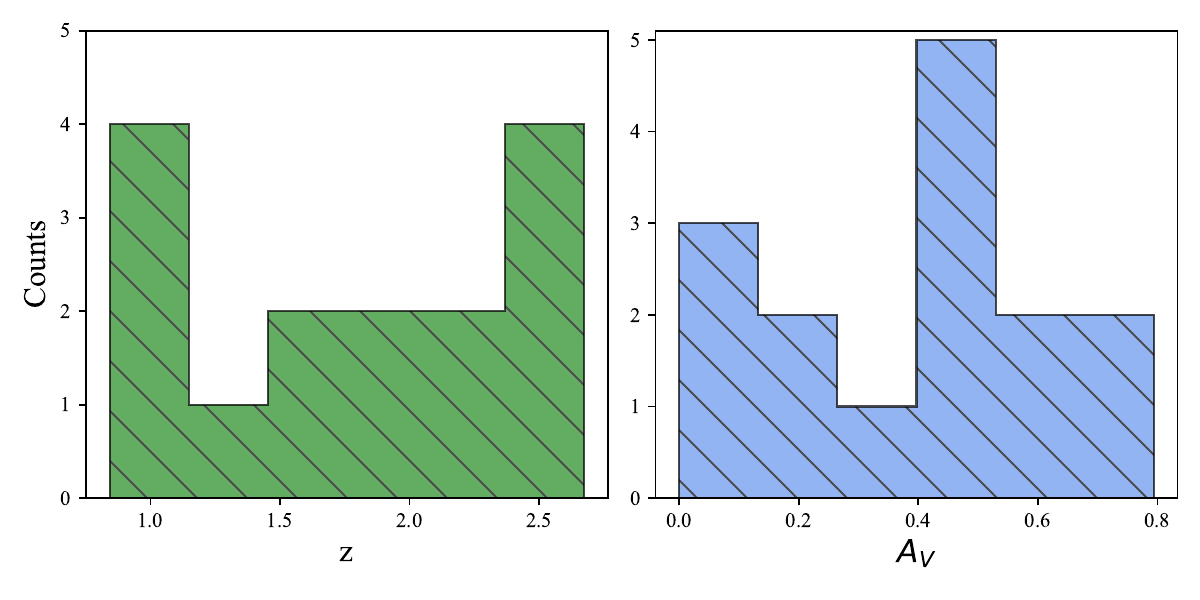}
    \caption{Histograms of redshift, $z$, and visual extinction parameters, $A_V$, for the parent quasar sample observed in this study.}
    \label{fig:Histograms}
\end{figure}

% Lensed quasar selection

Detecting lenses has been historically challenging, 
%with only $\sim 300$ known to date \citep{cambridge_lensed_quasars}, 
since it requires broad sky coverage and high angular resolution to separate the individual lensed images, as the angular separation is typically very small.
%and because most lensed quasars are very distant and thus faint. 
The spatial resolution of {\em Gaia} $\sim0.6\ \mathrm{arcsec}$ \citep{Gaia_2023}, compared to the $\sim 1-6\,\mathrm{arcsec}$ resolution of classical UV to infrared quasar survey \citep[e.g., SDSS, Pan-STARRS, 2MASS and AllWISE;][]{2002_SDSS,chambers2019panstarrs1surveys,2006_2MASS,2014_AllWise}, 
%and Pan-STARRS, $\sim1.2\ \mathrm{arcsec}$ \citep{chambers2019panstarrs1surveys}, $\sim2.5\ \mathrm{arcsec}$ for 2MASS \cite{2006_2MASS} and $\sim6\ \mathrm{arcsec}$ for AllWise \citep{2014_AllWise} 
is further optimal, particularly for identifying quasar lenses in close proximity.

We adopt the method of astrometric selection of particularly gravitationally lensed quasars by \citep{Gaia_2024} here, relying on their catalog of gravitationally lensed quasar candidates. Briefly, this was constructed in the following way: First, they combine major catalogs of quasars and AGN candidates \citep{Gaia_2023, Ducourant_2023, Flesch_2021, Shu_2019, Assef_2018}. Then, robust stellar sources were eliminated, with the following conditions: objects with proper motions larger than 14 mas/yr or parallaxes larger than 6 mas, along with objects with $G \geq 14 \ \mathrm{mag}$ or $(G_{BP} - G) > 1 \ \mathrm{mag}$ \& $(G - G_{RP}) > 0.8 \ \mathrm{mag}$ \& $G < 20 \ \mathrm{mag}$. To exclude local bright areas, e.g. H\,{\sc i} regions, sources in the direction of the small and large Magellanic clouds, major globular clusters or other galaxies \cite{Harris_2010} were also removed. All elimination criteria were conservatively chosen to maximize the chances of finding new lenses. Now, near all quasar candidates, other objects within 6" are detected using GravLens clustering algorithm \citep{Ester_1996}. The resulting catalog includes 76 lenses with four images (quads). Each source in the catalog is assigned likelihood of it being a lensed quasar, primarily based on the mean BP/RP spectra, as lensed images will have similar spectral shapes. We embarked on a spectroscopic follow-up campaign of the highest ranked quasar lens candidates (A- or higher, Heintz et al. in prep.), and have thus far secured the spectra of 11 sources. 

\begin{figure}[h]
    \centering
    \includegraphics[width=1\linewidth]{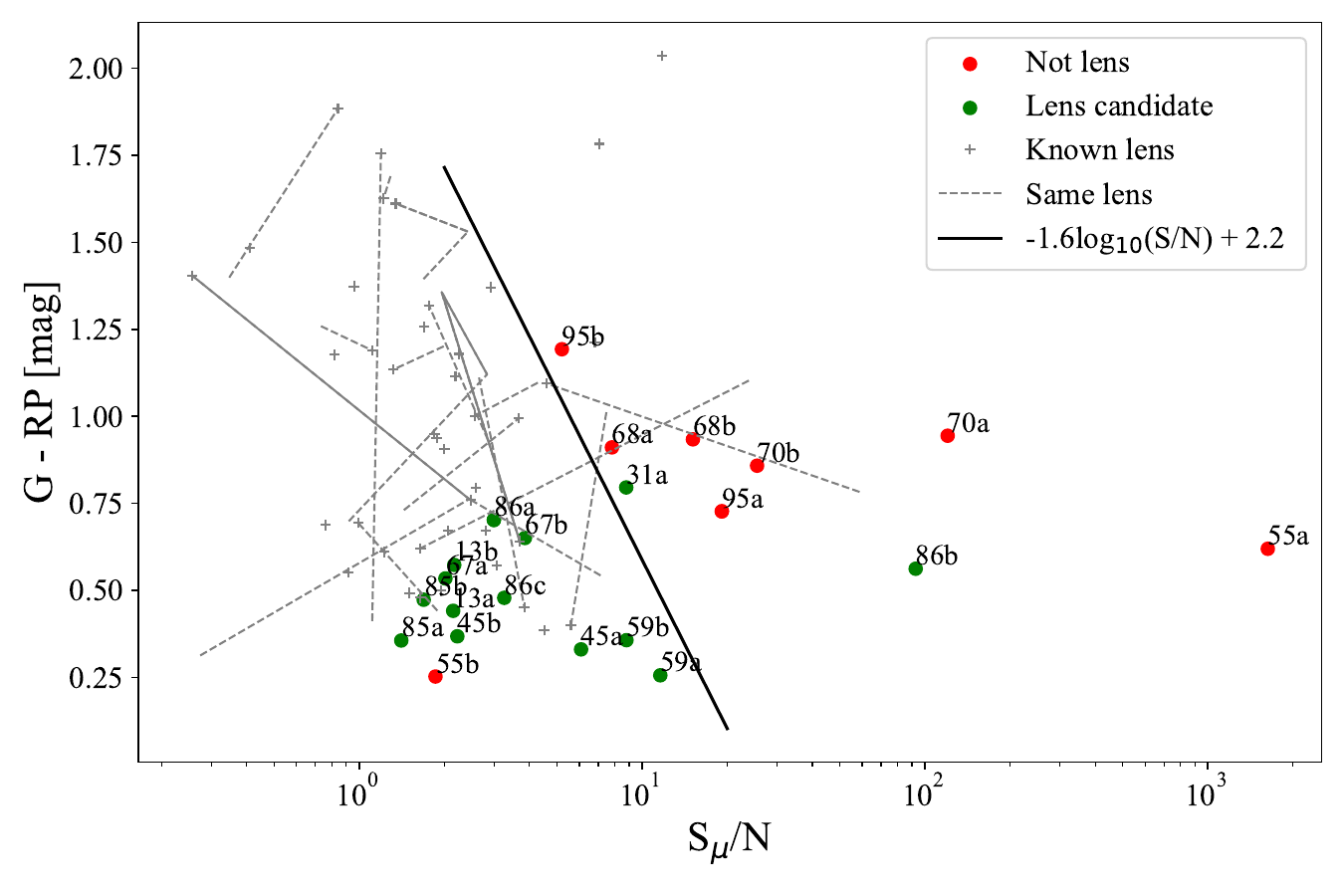}
    \caption{$S/N$ for proper motion against G - RP, the difference in the Gaia bands G- and RP-bands. The plotted points are sources from the study discarded as lens candidates (red) and remaining lens candidates, each identified with the source ID and an index separating individual sight lines in the source. To enhance the statistics, known lenses are added, either as lines to connect lenses where photometry of both objects exist in Gaia, or "+" if not. A line is added, showing that the lens and lens candidate population can be separated by an additional photometric criterion. Note, that we deselect only sources where all sight lines complies to the condition in Eq. \ref{eq:SelectionCondition} -- meaning they will be to the right of the line. }
    \label{fig:GaiaCriteria}
\end{figure}

% Summary paragraph
In this work, we report the independent discovery and characterization of a particularly unique system discovered as part of this spectroscopic campaign, the triple lensed quasar, named DR3Gaia2107-1611 at $z=2.67$. This was first discovered by \citet{2023_Dux}.
% and non-identical sight lines with variability in time. 
The paper is structured as follows. Sect.~\ref{sec:obs} reports the observational sample and the first results on the parent spectroscopy sample, based on which we propose an optimized quasar-lens selection.
%follow-up survey of 11 \textit{Gaia}-detected lensed quasar candidates and we propose an optimized quasar-lens selection. 
Sect.~\ref{sec:res} presents the results on the lens modelling of the unique system, DR3Gaia2107-1611, which is used to deconvolve each sight line from the observed time to the intrinsic quasar time. The emission-line properties are then tracked over $\sim18\mathrm{months}$ at a nearly monthly cadence in the quasar rest-frame. In Sect.~\ref{sec:disc}, we discuss the possibility of microlensing in one of the sight lines, as well as the interpretation of the observed temporal variation of the broad emission lines. We conclude on our work in Sect.~\ref{sec:conc}.   

Throughout the paper, we assume the flat $\Lambda$CDM-dominated concordance cosmological model \citep{2020_Planck} used to constrain the lens model, and report magnitudes in the AB system. 

\begin{figure*}
    \centering
    \includegraphics[width=1\linewidth]{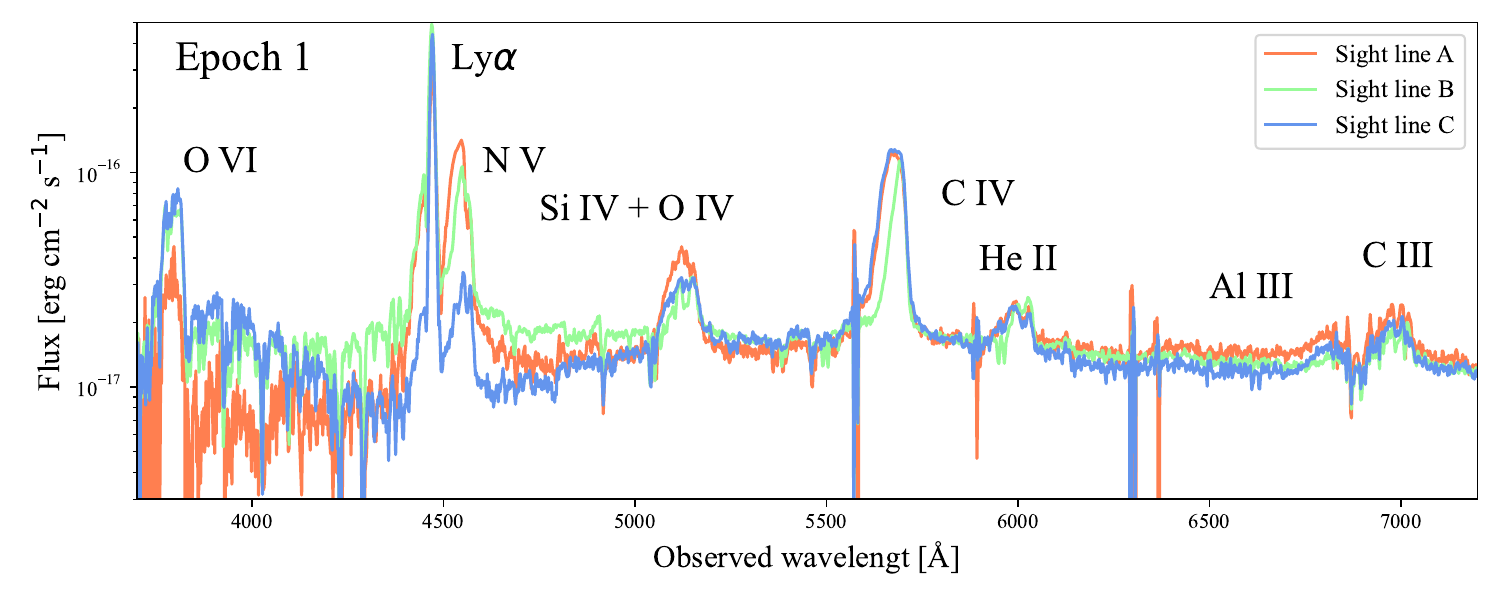}
    \caption{(Left): Spectra of each sight line in the lensed quasar DR3Gaia2107-1611, as seen in Fig. \ref{fig:DR3Gaia}. The variability between sightlines are especially clear in the shape of C\,IV, which is different for sight line B, while Ly$\alpha$ vary substantially across all sightlines.}
    \label{fig:epoch1}
\end{figure*}

\begin{figure}
    \centering
    \includegraphics[width=1\linewidth]{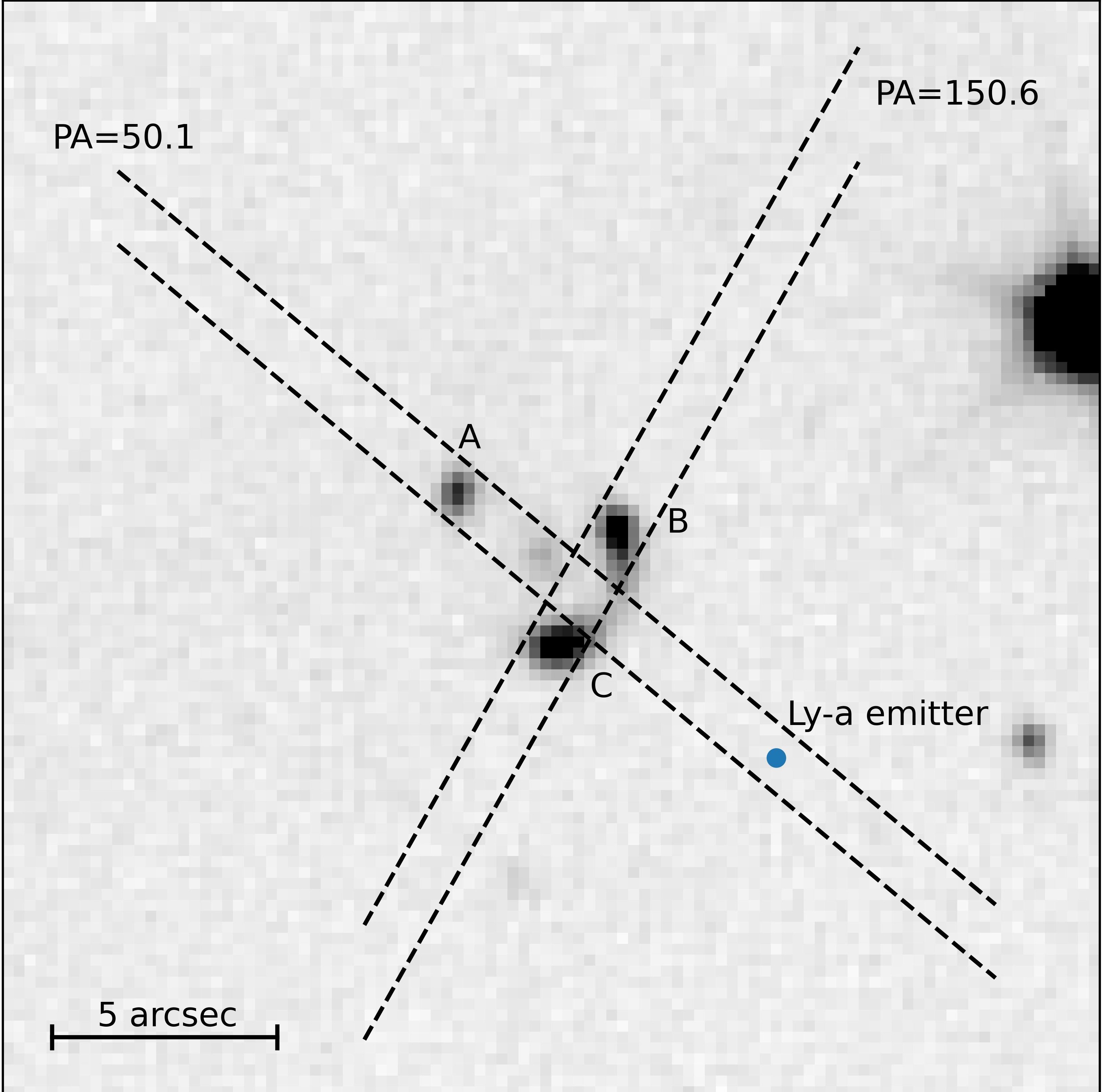}
    \caption{Imaging of DR3Gaia2107-1611 with each slit used for the observing the three sightlines, A, B and C, as seen in Fig. \ref{fig:epoch1}. The lens galaxy is vaguely visible in the center of the lens system. We mark the tentative detection of Ly$\alpha$ emission on the opposite side of sight line A, detected from the 2D-spectrum of the system before extraction. }
    \label{fig:DR3Gaia}
\end{figure}

\section{Observations and sample analysis}
\label{sec:obs}
Our spectroscopic campaign of the highest ranked lensed quasar candidates from the \cite{Gaia_2024} was carried out with the Gran Telescopio Canarias (GTC; see Table~\ref{tab:Observational_details} for details). For the remainder of this study, we report only the first four digits of each coordinate to identify the candidate-lens object. The observations are acquired using the Optical System for Imaging and low-Intermediate-Resolution Integrated Spectroscopy (OSIRIS) instrument and the R1000B grism \citep{gtc_osiris} mounted at GTC. 

\begin{table}[b]
    \centering
    \begin{tabular}{cc}
         Object & Classification  \\
         \hline
         DR3Gaia055409.442-234754.13 & Candidate \\
         \hline
         DR3Gaia210752-161131 & Lens \\
         \hline
         DR3Gaia050613.596-253047.45 & Lens \\
         \hline
         DR3Gaia111221.158-201111.55 & Proj.$^ {\alpha}$ QSO + galaxy \\
         \hline
         DR3Gaia020501.994-323348.59 & Proj. QSO+QSO \\
         \hline
         DR3Gaia045755.331+124238.67 & Proj. QSO+star+galaxy \\
         \hline
         DR3Gaia070020.352+132813.68 & Proj. star+star \\
         \hline
         DR3Gaia044652.260-310219.85 & Candidate \\
         \hline
         DR3Gaia115352.588-252027.70 & Proj. QSO+star \\
         \hline
         DR3Gaia031013.747+352414.86 & Proj. galaxy+star \\
         \hline
         DR3Gaia114934.110-172651.95 & Proj. QSO+star \\
         \hline
    \end{tabular}
    \caption{Overview of the classification of all 11 objects studied as a part of a larger follow-up spectroscopy survey of lensed quasar candidates. All objects are classified by identifying each sight line of the source. All objects will be named using just the 4 first digits of each coordinate. $^\alpha$Proj. refers to projected.}
    \label{tab:Observational_details}
\end{table}

\begin{figure*}[t]
    \centering
    \includegraphics[width=1\linewidth]{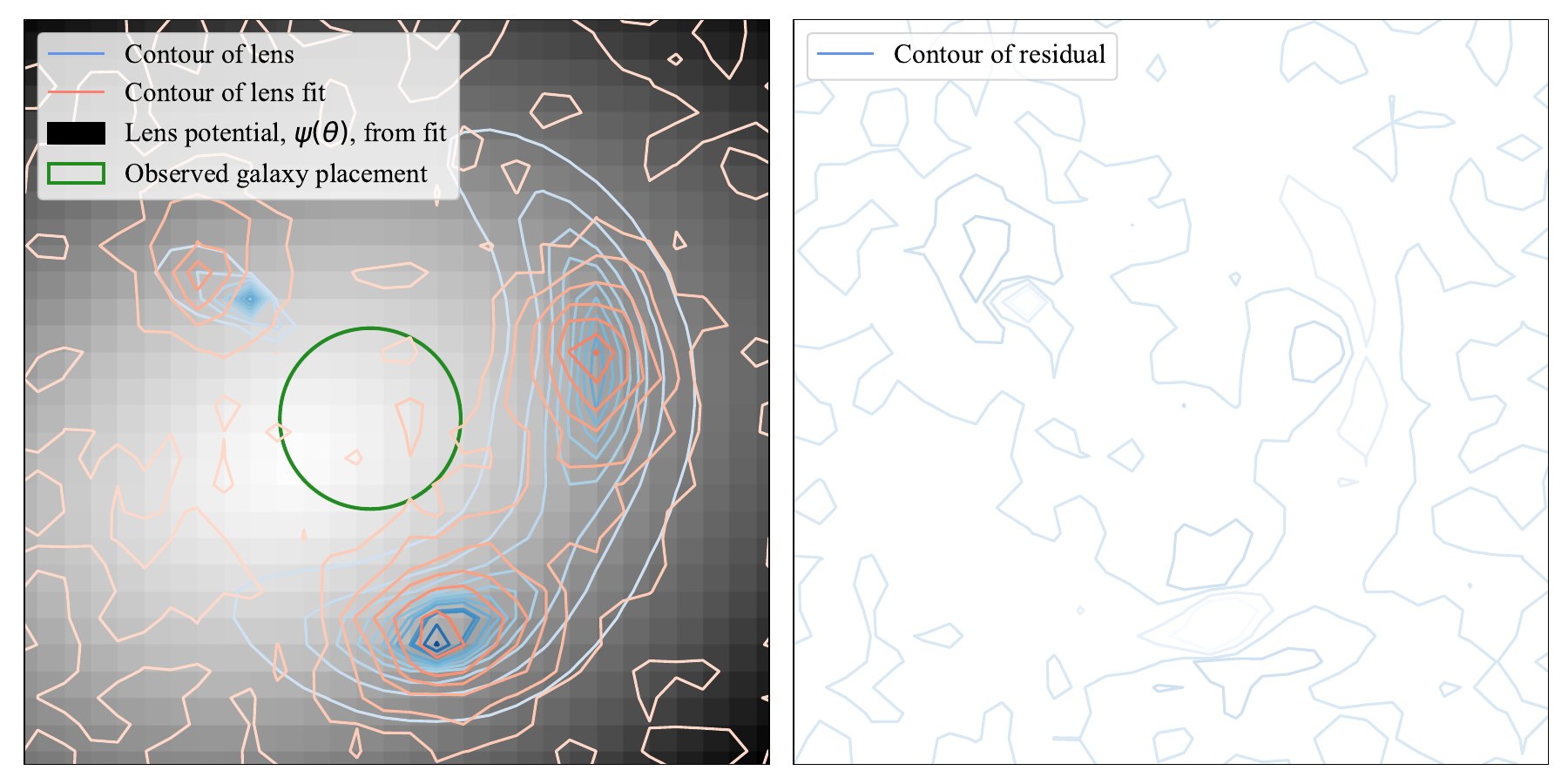}
    \caption{Comparison of the result of the fit of the lens configuration to the data (left) with the overlay of the contours of the data and the fit, as well as an illustration of the potential $\psi(\theta)$ of the lens galaxy from the fit. (Right) the residual of the data and the fit with the same contour levels as the fit and data. The residual reflects predominantly statistical noise.  
    The fit first and foremost allows us to determine the ordering of sightlines. This allow us to translate spectrum properties into the intrinsic time of the quasar. A discrepancy is visible in the comparison between lens and lens model (left), which likely stems from the fact that the top left sight line is a significantly smaller signal than the remaining. However, we note that the deviation of the location of the centre is $\delta_{RA}\sim0.4"$ and $\delta_{DEC}\sim0.2"$, which is small enough that we expect it not to influence the derived properties.}
    \label{fig:Lens_fit}
\end{figure*}

 \begin{figure}
    \centering
    \includegraphics[width=1\linewidth]{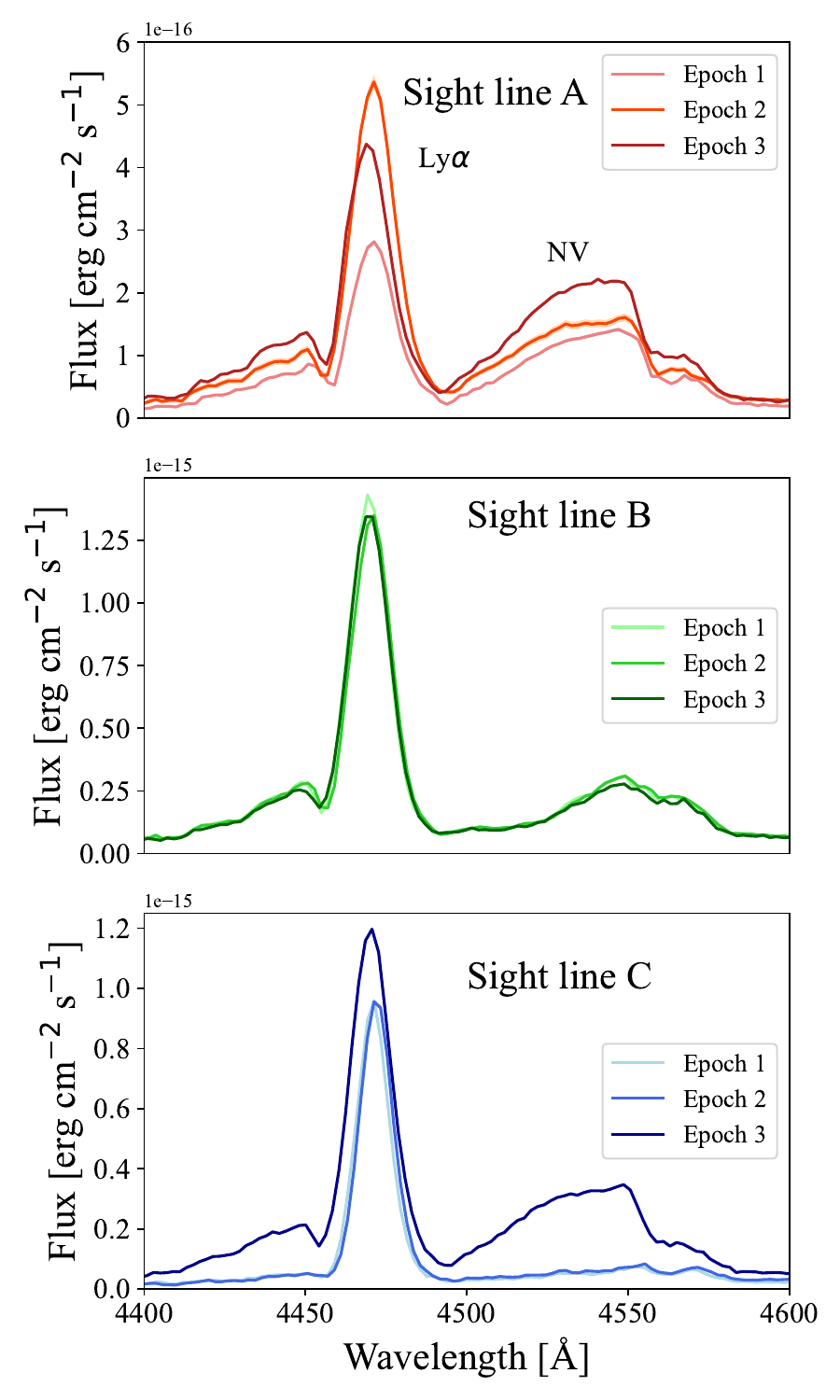}
    \caption{The Ly$\alpha$ emission line of each sight line of the lensed quasar DR3Gaia2107-1611 across three different epochs, as described in Table \ref{tab:epoch_details}. Ly$\alpha$ changes dramatically along sightline A, modestly in sightline C and mildly along sightline B. We show only the Ly$\alpha$ line here to highlight the observed variation. }
    \label{fig:all_epochs}
\end{figure}

 \begin{figure*}
     \centering
     \includegraphics[width=1\linewidth]{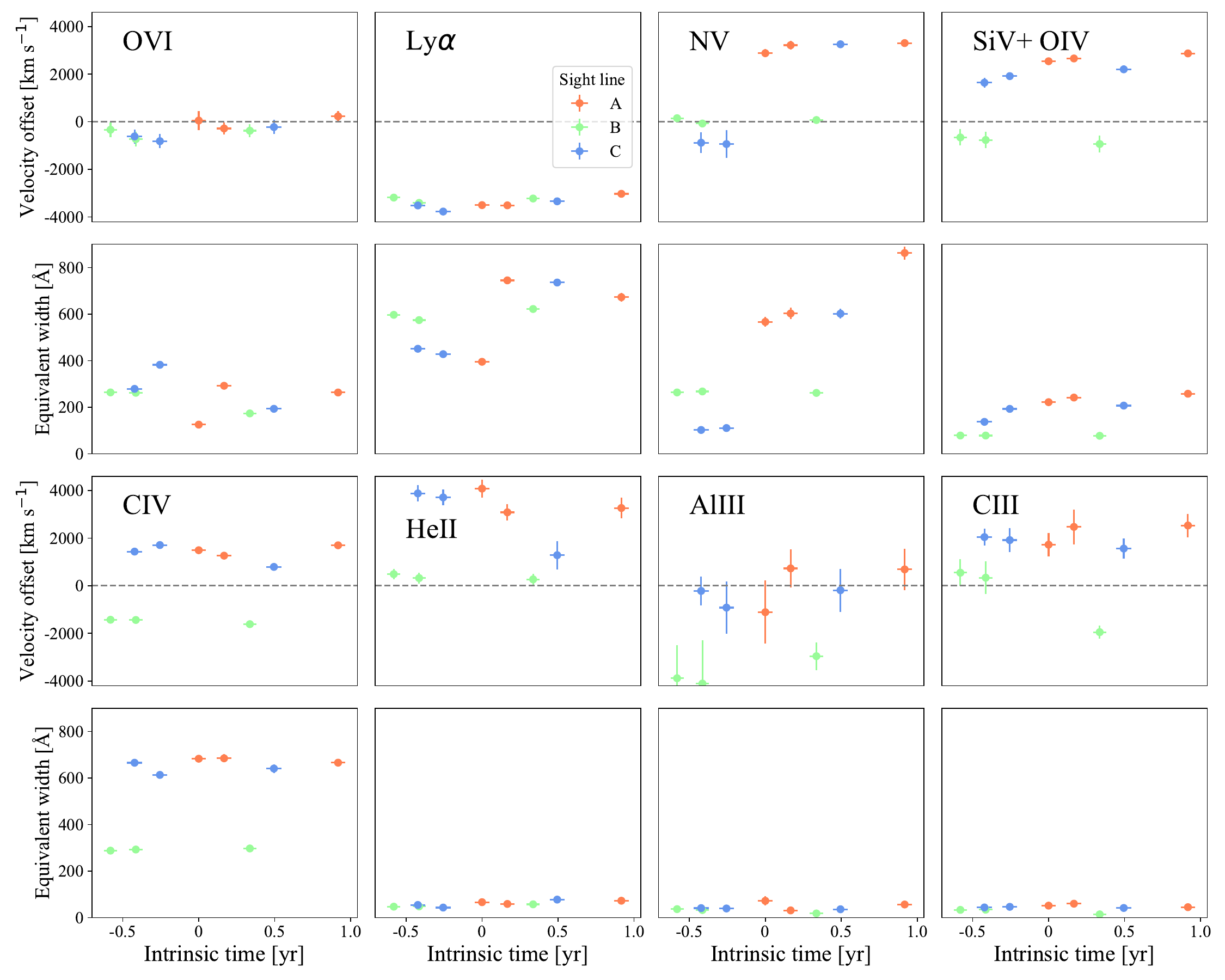}
     \caption{Change in velocity offset (upper row) and equivalent width (EW) (lower row) for all emission lines of each sight line from epoch to epoch. The intrinsic quasar time is deconvolved to the intrinsic quasar time using the time delays determined from the lens model (see Table \ref{tab:time_delay}).     
     The collective change in the velocity offset indicates that the change of the emission lines are related to a rotation of the quasar or a change in strength of outflow of material from the quasar.}
     \label{fig:emission_changes}
 \end{figure*}

For each target, we attempt to align the slit to cover at minimum two of the potential lensed images. The spectra are reduced and extracted using the PyLongslit reduction pipeline \citep{Kostas_2025}. As the different spectral object traces, the foreground lens and the background lensed quasars, are placed in very close proximity to each other on the detector, we extract the spectra using another approach than the commonly used optimal-extraction method. We trace the separate objects first by hand-tracing around a hundred points of where the object is placed, and then fitting a polynomial model to these points. In the extraction, we define a fixed region of suitable amount of pixels around the polynomial trace, and then extract by summing detector counts in this region using the photutils Python package \citep{photutils}. There are several reasons to motivate this approach. Firstly, mathematically modelling the object traces with full automation fails due to the close proximity of the neighbouring traces. Secondly, the different spectra overlap in the bright emission regions. Defining a fixed region around the trace allows us to avoid including flux from neighbouring objects in the extraction. This, however, does result in some flux loss and potentially higher noise level, as the fixed region both over and under-estimates the object extent depending on the spectral region. 

\begin{table}[b]
    \centering
    \begin{tabular}{c|cc}
          & Date &Telescope, Instrument \\
         \hline
         Epoch 1 & 07/08/2024 & GTC, OSIRIS \\
         \hline
         Epoch 2 & 05/10/2024 & GTC, OSIRIS \\
         \hline
         Epoch 3 & 20/07/2025 & GTC, OSIRIS \\
         \hline
    \end{tabular}
    \caption{Observational details of epoch 1, 2 and 3. All epochs are taken using the OSIRIS instrument on the GTC.}
    \label{tab:epoch_details}
\end{table}

\begin{table}[b]
    \centering
    \begin{tabular}{c|c}
    \hline
    \multicolumn{2}{c}{Lens galaxy, $z=0.63$} \\
    \hline
    Centre 1 & $-14.35"^{+0.04}_{-0.02}$ \\
    Centre 2 & $ -14.46"^{+0.05}_{-0.03}$ \\
    Elliptical component 1 & $-0.317"\pm 0.01" $ \\
    Elliptical component 2 & $0.28"\pm0.01"$ \\
    Einstein radius & $2.35" \pm0.02"$ \\

    \hline
    \multicolumn{2}{c}{Source quasar, $z=2.67$} \\
    \hline
    Centre 1 & $ -15.03"\pm0.01"$ \\
    Centre 2 & $-13.50"^{+0.05"}_{-0.01"}$ \\
    Elliptical component 1 & $-0.20"^{+0.03"}_{-0.02"}$ \\
    Elliptical component 2 & $0.09"^{+0.01"}_{-0.04"}$ \\
    Effective radius & $0.77"\pm0.08"$ \\
    Sersic index & $2.44"^{+0.02"}_{-0.05"}$ \\
    \hline
         
    \end{tabular}
    \caption{Fit parameters determined from the fit of the lens system DR3Gaia2107-1611. The lens galaxy is modelled as an isothermal mass profile with no luminosity for simplicity, while the quasar is modelled as point mass with a sersic luminosity profile. The elliptical components refer to its foci relative to the centre, which determine the shape of the ellipse \citep{Nightingale_2021}.}
    \label{tab:fit_parameters}
\end{table}

\begin{table}[t]
    \centering
    \begin{tabular}{cc}
    \hline
    $\Delta\tau_{A,B}$ & (-209.0$\pm$ 17) days \\
    $\Delta\tau_{A,C}$ & (-151.7$\pm$16 ) days \\
    $\Delta\tau_{B,C}$ & (57.8$\pm$18) days \\
    \hline
    \end{tabular}
    \caption{Time delay between sight lines, $\Delta\tau_{X,Y} = \tau_Y-\tau_X$, as found from Eq. \ref{eq:time_delay} using the potential found from the mass profile of the lens galaxy of the fit of the system (see Table \ref{tab:fit_parameters}) and the positions of each sight line, $\vec{\theta}$. Thus, a positive time corresponding to the second index lacking behind the first index, and the other way around. Thus, the timely order of the sight lines are; B, C, A.}
    \label{tab:time_delay}
\end{table}

\subsection{Spectroscopic sample completeness}

All 11 target systems are individually examined to characterize their origin. Stellar objects are determined based on a lack of strong emission lines and through a fit to a stellar spectrum using Pyhammer \citep{2020_Pyhammer}, determining also the stellar type. Quasars (and galaxies) are determined by their prominent broad (or narrow) emission and/or absorption lines, as well as the continuum profile. 
The majority of the target systems are either projected quasar pairs or quasar/galaxy + star pairs, with a total of 16 unique quasars confirmed.
The redshift, $z$, is determined for all galaxies and quasars based on the identified emission, and the visual dust extinction parameter, $A_V$, is determined for all quasars by fitting a dust-reddened quasar template \citep[][see Fig. \ref{fig:dust_fit}]{selsing16}, with the extinction curve described by

\begin{equation}
    \xi(\lambda) = \sum_{i=1}^{6} \frac{a_i}{(\lambda/\lambda_i)^{n_i} + (\lambda_i/\lambda)^{n_i} + b_i},
    \label{eq:dust_model}
\end{equation}

from \citet{1992_Pei} (see Fig \ref{fig:dust_fit} for an example). We assume the model parameters found for the Small Magellanic Cloud (SMC).
The distribution of redshift, $z$, and visual dust extinction parameters, $A_V$, for all quasars is shown in Figure \ref{fig:Histograms}. 
This shows a relatively broad distribution of both redshift and visual dust extinction, ranging from $z\sim 1-2.5$ and $A_V \sim 0 - 1$\,mag. 
% Specifically, the redshift distribution mirrors the general quasar population, as a decrease of quasars are found around $z\sim3$ (\cite{2010_Schneider}, figure 5). Furthermore, with $A_V$'s spanning almost from 0-1.0, suggests a wide range of dust along the line of sight \citep{2015_Zafar}, and thus no bias in deselecting dusty quasars is indicated. This suggest that there is indeed no colour bias in the astrometrically selected lensed quasar candidates. These statistics is despite the fact that the sampling size is incredible small, with only 16 quasars are detected in this study. Furthermore, the the measure of redshift and dust extinction parameters cannot be assumed to be entirely independent measurements, since two or more quasars are often studied very close together. Specifically for the case of lensed quasars, the redshift is just different observations of the same objects. Thus, the broad range of values are the more statistically impressive.

To determine which objects are lenses, the following procedure is used: If only one or less source within the object is a quasar, it is immediately discarded as a lens candidate. If a lens galaxy is identified either in imaging or the spectrum of both sources, the object is verified as a lens. For the remaining objects, a probability model is constructed based on assuming the objects to be either quasar pairs or lensed quasars, and assuming the population of quasar mergers to be negligible \citep[e.g.,][]{Schawinski_2012}. 
%Now, quasar pairs are typically associated with a larger sky separation than lensed quasars. However, the sky separation of lensed quasars depends on the mass of the lens, the redshift of the quasar and the alignment, and can thus vary dramatically, e.g. the Sloan Digital Sky Survey (SDSS) discovered a lensed quasar with a sky separation of 14.62" \citep{Inada_2003}. 
%Thus, the sky separation alone is not a clear way to exclude sources as gravitational lenses. However, the probability for larger separations for lensed quasars grows with higher redshift. 
The sky separation and the redshift of known lenses and known quasar pairs can be used to determine a probability for whether or not the system is a lens,

\begin{align}
    P(\mathrm{lens}|z,\theta) = \frac{  P(z,\theta |\mathrm{lens}) P(\mathrm{lensed})}{P(z,\theta)} 
    \label{eq:Bayes}
\end{align}
Where $P(z,\theta)=P(z,\theta|\mathrm{lens})P(\mathrm{lens})+P(z,\theta|\mathrm{\neg\ lens}) P(\mathrm{\neg lens})$. The probability of $z$ and $\theta$ given that a source is known to be a lens, $P(z,\theta|\mathrm{lens})$, is constructed observationally by using the Cambridge QSO lens database of 220 known lensed QSOs \citep{cambridge_lensed_quasars}, out of which $210$ has determined redshift and separation. $P(z,\theta|\mathrm{\neg\ lens})$ is constructed using a catalogue of 1842 quasar pair candidates, where $\sim900$ is characterized as quasar pairs \citep{jing2025quasarpaircatalogcompiled}.
The prior $P(\mathrm{lens})$, the probability of a double quasar being a lens instead of a quasar pair, is estimated assuming the two catalogs to be complete in terms of the ratio between lenses and pairs, and thus $P(\mathrm{lens})= 0.195$, determined on the base of the 220 lensed QSOs in \cite{cambridge_lensed_quasars} and 907 QSO pairs in \cite{jing2025quasarpaircatalogcompiled}. Assuming the number of quasar mergers to be negligible, this yields $P(\mathrm{\neg lens}) = 1-P(\mathrm{lens})$.

As a result, DR3Gaia0506-2530 and DR3Gaia2107-1611 are determined to be lensed quasars based on the visibility of the lens galaxy, while DR3Gaia0446-3102 and DR3Gaia0554-2347 keep their lens candidate status. The remaining objects are discarded, primarily based on only one or less quasar source being robustly identified in the systems, while only two are removed due to their probability of being a lens being $>2\sigma$.

\subsection{An optimized astrometric selection}
\label{sec:OwnSelection}

Considering now the full parent spectroscopic sample of candidate lens quasars, we determine an additional photometric selection criteria, to reduce the amount of contamination in the selection. 
%To make sure not to deselect any actual lens systems, one would wish to select all double (or more) quasars. Thus the selection criterium should should preferably only remove the sources with only one quasar sight line. We note, that it is not directly applicable to use the probability estimate developed above, since the redshift is generally not known from photometry. Furthermore, we note that among the large sky surveys, Gaia has the best resolution, in contrast to e.g. 2MASS, AllWise. In fact, for almost all the studied sources here, only Gaia completely resolves all objects. This motivates Gaia criteria as superior, and we will focus on this. 
Along with the objects in this study, the sample of lensed quasars are enhanced by taking $\sim$100 known lensed quasars \citep{cambridge_lensed_quasars}. Using an explorative approach among the {\em Gaia} photometric and astrometric measurements, we highlight the following demarcation line for an optimized selection of lensed quasars, also requiring at least two astrometrically-identified quasar candidates in close on-sky proximity;

\begin{equation}
    G - RP < -1.6\log_{10}(S_{\mu}/N) + 2.2,
    \label{eq:SelectionCondition}
\end{equation}

\noindent where $S_\mu/N$ is the signal-to-noise ratio of the proper motion, see also Figure \ref{fig:GaiaCriteria}.

\section{Results on D3Gaia2107-1611}
\label{sec:res}
\begin{figure}
    \centering
    \includegraphics[width=1\linewidth]{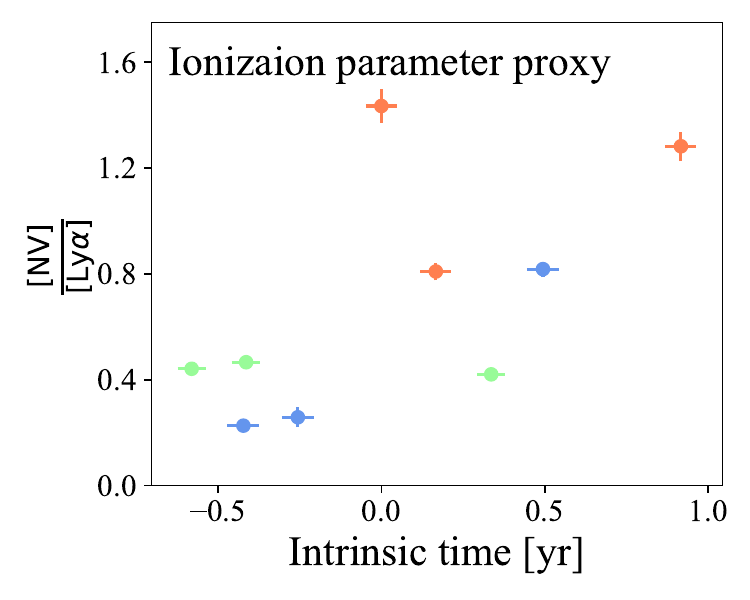}
    \caption{ Proxies for the  ionization parameter (see Eq. \ref{eq:ionization}) as a function of the intrinsic time of the quasar. }
    \label{fig:met_and_ion_parameters}
\end{figure}

The source DR3Gaia2107-1611 is determined to be a lensed quasar based on three sight lines consisting of quasars with identical redshift, and the lens galaxy being clearly visible in both the spectrum and imaging in the $i$- and $z$-bands \citep{PanSTARRS}.  
This prompted our follow-up spectroscopic campaign of the source, as well as the construction of a lens model of the system, critical in order to present, analyze and interpret the spectroscopic results of a lens system.

We note that this object was first discovered by \citet{2023_Dux}, but has been independently detected as a part of this study. In this work, we characterize it over 3 epochs.

\subsection{Lens model of DR3Gaia2107-1611} 
\label{sec:LensModel}
Based on the imaging data from PanSTARRS \citep{PanSTARRS} in the $i$-band along with the redshift of the lens and the source, a lens model is fitted to the object, providing the geometry of the system using PyAutolens \citep{Nightingale_2021}. The fit assumes an isothermal mass distribution for the lens galaxy and a point mass distribution for the quasar with a Sérsic luminosity profile. The resulting fit, including the lens potential, is seen in Figure \ref{fig:Lens_fit}, along with the residual plotted on the same colormap scale. The residual is mostly random, indicating a good match of the fitted model to the data. We note, however, a systematic deviation in the placement of sightline A, as well as the placement of the lens galaxy. This is likely due to sightline B and C dominating the signal, and thus the small luminosity being less constrained in the fit. We therefore expect the effect to be negligible, though it could lead to an underestimation of the time difference in the calculations using this sightline. 

The construction of the lens model through the fit allows us to determine the time delay between the sight lines. This enable a deconvolution of the observed time into an intrinsic time-frame of the quasar, and thus translate spectrum properties into this time. The best-fit parameters are summarized in Table \ref{tab:fit_parameters}. The gravitational potential found from the fit is used to find the time delay between each line of sight,

\begin{equation}
    t(\vec{\theta}) = \frac{1+z_{lens}}{c}\frac{D_{source}D_{lens}}{D_{lens,source}}\left[ \frac{1}{2}(\vec{\theta}-\vec{\alpha}) - \psi(\vec{\theta}) \right],
    \label{eq:time_delay}
\end{equation}

where $D_{source}$, $D_{lens}$ and $D_{lens,source}$ are angular diameter distances, $\vec{\theta}$ and $\vec{\alpha}$ are the angle from an optic reference point to the apparent source position and actual source position, respectively. $\psi(\vec{\theta})$ is the potential value at the point of the apparent position, $\theta$. This gives the time delays between the three lines of sight (see Table \ref{tab:time_delay}). Thus, the chronological order of the sightlines is B, C, A. The time delay is used to convert the time change of each emission line into the intrinsic time of the quasar (see Figure \ref{fig:emission_changes}). %The maximal possible scale of change between the sight lines is $\sim c\tau_0$.
From the determined Einstein radius, $\theta_E$, we can further determine the halo mass of the lens galaxy using the Einstein lens equation, yielding $M_{\rm halo} = (2.78 \pm 0.05)\times 10^{10}\,M_\odot$.

\subsection{The spectrum of DR3Gaia2107-1611 }
% Possibilities: variability, microlensing, absorption, not the same object

The spectrum of epoch 1 of the lensed system is shown in Fig.~\ref{fig:epoch1}, along with the imaging with indrawn slits used for observing in Fig.~\ref{fig:DR3Gaia}.
The redshift of the quasar is determined from the identified emission lines to be $z=2.67\pm0.03$. Since the galaxy spectrum can be extracted from the 2D longslit spectrum as well, the redshift of the lens galaxy is determined as $z=0.630\pm0.001$ based on the CaH, K and G absorption lines from the lens galaxy spectrum. 
We obtained follow-up spectroscopy on 5th of October 2024 and 20th of July 2025 (see Table \ref{tab:epoch_details}) to look for potential variability of the source. To highlight the observed variability, the Ly$\alpha$ of each spectrum obtained at each epoch for each sightline is shown in Figure~\ref{fig:all_epochs}.

For epoch 1, this source shows remarkable variations in the emission line profiles of each sight line (see Figure \ref{fig:epoch1}), where especially the Ly$\alpha$, S\,{\sc iv}+O\,{\sc iv} and C\,{\sc iv} emission lines clearly vary from sightline to sightline.

When considering each sightline across different epochs, the variation is quantified by two basic properties of emission lines, the velocity offset, $\Delta v = (z_{sys}-z_{line})c$, and the equivalent width (EW), which for constant continuum is $W_\lambda = A/F_c$, where $A$ is the flux of any given emission line and $F_c$ is the constant value of the continuum. We caution that the definition of $z_{sys}$ is hard to define for this specific object, since the emission lines change between sight lines and between epochs. In this work, we simply define $z_{sys}$ as the median redshift measured from each spectrum and across epochs . This slight ambiguity makes it difficult to interpret $\Delta v$ directly as outflow from the quasar in a direction towards/away from the observer. Instead, we simply use $\Delta v$ as a relative difference of the flow velocity. To create a time series of $\Delta v$ and $W_\lambda$, the time scale of each sight line is converted into an intrinsic time of the quasar by offsetting the observations from one another using the time difference in Table \ref{tab:time_delay} found from the lens model (see Section \ref{sec:LensModel}). The result is plotted in Figure \ref{fig:emission_changes}. 

Firstly, we remark that sightline B does not follow the general trend of sightlines A and C for most of the emission lines (this is especially clear in the N\,{\sc v}, S\,{\sc iv}+O\,{\sc iv}, and C\,{\sc iv} emission lines (see Figure \ref{fig:emission_changes})). This discrepant behaviour must either be due to the lens model not capturing the actual behaviour of the quasar however, this is believed to be highly unlikely, due to sight line B and C being on symmetric sides of the caustic. This is evaluated further in Sect \ref{sec:disc:lens_model}. Alternatively, sightline B is subjected to microlensing by a local overdensity in the foreground galaxy. This is evaluated in the discussion, Section \ref{sec:disc:lens_model}.

Considering only sightline A and C, a temporal increase in both velocity offset and equivalent width (EW) is observed for the N\,{\sc v}, S\,{\sc iv}+O\,{\sc iv}, and C\,{\sc iv} emission lines. A temporal increase of the EW is observed for Ly$\alpha$ emission. Lastly, a temporal change in the velocity offset, He\,{\sc ii}, Al\,{\sc iii}, and C\,{\sc iii} emissions are indicated. %The change of EW can be translated to a change of metallicity by using the weakly temperature dependent metallicity proxy \citep{Hamann_2002},
%\begin{equation}
%    \frac{Z}{Z_\odot} \propto  \frac{[\mathrm{NV} \lambda 1240]}{[\mathrm{CIV}\lambda1549]+[\mathrm{OVI\lambda1033 }]}  ,
%    \label{eq:metallicity}
%\end{equation}
%where $[\mathrm{X}\lambda \mathrm{Y}]$ refer to the EW of emission $X$ at wavelength $Y$. 
The change of EW can be translated to a change of the ionization parameter, by use of the proxy determined in \citet{Temple_2021} to be the diagnostically robust,

\begin{equation}
    U = \frac{\phi(H)}{cn_H} \propto \frac{[\mathrm{NV}\lambda1240]}{[\mathrm{Ly}\alpha\lambda 1215]}.
    \label{eq:ionization}
\end{equation}

where $[\mathrm{X}\lambda \mathrm{Y}]$ refer to the EW of emission $X$ at wavelength $Y$. We note that an increase in this parameter can also be explained by a change of other physical parameters, such as the microturbulence parameter, $\nu_{turb}$ \citep{Temple_2021}, describing the expected turbulent motion in the outflow winds and accretion disks. %However, we note that the ionization is expected to increase with increasing metallicity, and since the two proxies increase together, the ionization parameters is believed to be a reasonable proxy for ionization alone in this case.
The ionization as a function of time is shown in Figure \ref{fig:met_and_ion_parameters}. 

\section{Discussion}
\label{sec:disc}
% Er det lens effekt eller faktisk fysisk ændring?
% Forudsigelse udfra fortsat tidslig data?

\subsection{Microlensing or inaccurate lens model?}
\label{sec:disc:lens_model}

Since we know that all sightlines originate from the same quasar, the different behaviour of sightline B compared to sightlines A and C can be interpreted as either; the lens model not being precise enough to describe the system or alternatively, that sightline B is subject to microlensing or other local variations in the foreground lens galaxy.

The microlensing interpretation is in agreement with the variations seen in the EW of the C\,{\em iv} emission line between sightline B and sightlines A and C, since this feature likely probes the quasar accretion disk winds \citep{2012_Richards}. Consequently, sightline B may originate from a part of the quasar with weaker accretion disk winds.

The inaccurate lens model interpretation is consistent with the qualitative deviations between the lens model and the lens system, specifically the deviation of the lens position 
$\delta_{RA}\sim0.4^"$ and $\delta_{DEC}\sim0.2^"$  and the position of sightline A is $\delta_{RA}\sim0.2^"$ and $\delta_{DEC}\sim0.1^"$. The deviation is systematic but relatively small compared to the scale of the object $\sim4^"$. For the lens model in this work, the time difference for sight line B is $\Delta\tau_{A,B}=(-209\pm17)\mathrm{days}$, and $\Delta\tau_{B,C}=(57.8\pm18)\mathrm{days}$. For an inaccuracy in the model to correct the discrepancy between sightline B and the remaining sight lines, this time delay would have to increase by at least a factor of $\sim1\mathrm{yr}$, resulting in the largest time delay increasing to almost $\Delta\tau_{A,B}\sim-2\mathrm{yr}$. The time delays of lenses are typically within the range of  $\sim(10^{-2}-10^2)\mathrm{days}$
\citep{Oguri_2010, Eulaers_2011}. However, it is not unphysical with a time delay of the order $\sim2\mathrm{yr}$, though it would be extraordinarily high. Furthermore, this should only affect sightline B. But sightlines B and C are positioned almost symmetrically in the lens configuration, and this discrepancy would only be possible if the mass distribution of the galaxy is highly inhomogeneous in areas large enough to significantly alter the gravitational potential to be asymmetric. This would also explain the inaccuracy of the lens model, since the galaxy mass distribution is assumed to be isothermal. Old elliptical galaxies, common among lenses, typically show spherically symmetric density distributions \citep[e.g.][]{Chae_2013}.

Essentially, the case of interpreting the nature of sightline B, is either due to a large-scale over density of the lens galaxy, resulting in an asymmetric lens potential and an inaccurate lens model, or alternatively a small over density in the specific area of the light path for sightline B, resulting in a microlensing phenomena. The latter is believed to be the most plausible. We also note, that a test of the lens model is possible with further observations, as the inaccurate lens model interpretation specifically predict sightline B to catch up on the behaviour of sight line A and C. 

\subsection{Intrinsic change of quasar or changing sight line?}
\label{sec:disc:QSO_change}
The observed temporal change of the broad emission lines of the quasar can be explained either by a sudden and continued temporal change of the ionizing continuum due to, for example accretion rate fluctuations, or alternatively, resulting from a change in the part of the quasar sampled along our line of sight, with a more ionized region moving into view.
 
Both of these interpretations are consistent with the simultaneously increasing outflow of heavier elements (N\,{\sc v}, He\,{sc ii}, Al\,{\sc iii}, and C\,{\sc iii}), which correlates strongly with outflow in quasars \citep{Wang_2012}.
The velocity distribution of C\,{\sc iv}, measured by the distribution of FWHM of the EW for a large sample of quasars in \cite{Coatman_2016}, yields $\mathrm{FWHM_{EW,CIV}}\sim50 \mathrm{Å}$, similar to the variations in EW of C\,{\sc iv} as observed in DR3Gaia2107-1611 (see Figure \ref{fig:emission_changes}). The change of the velocity offset of C\,{\sc iv} is consistent with a slight change of what part of the quasar we observe.
Meanwhile, the time scale of changing-look AGNs are of the order $\sim\mathrm{months-years}$ \citep[][e.g.]{2015_LaMassa, 2015_Runnoe, MacLeod_2016}, consistent with the temporal variance in the quasar-lens system studied here.
The present observations are thus inconclusive in the likely origin of the observed variability in the source, likely requiring a more comprehensive follow-up campaign or more detailed lens modeling.
% change is due to an intrinsic quasar change or a simply what part of the quasar dominate along our line of sight.

% Interestingly, we note a sudden spike in both metallicity and ionization, which decrease, before it continues to follow an increasing trend. This could indicate that the ionization is rapidly onset, after which is continues to increase continuously.

\section{Conclusions}
\label{sec:conc}
In this work, we presented the results of a spectroscopic campaign, currently including 11 gravitationally lensed quasar candidates selected purely based on measurements from the \cite{Gaia_2024}. We summarized the sample completeness and redshift distribution of the 16 uniquely identified quasars. Three of the candidate 11 systems were robustly determined as lensed quasars, and an additional system remain as a lens candidate only. We used our first pilot observations to define a new joint astrometric+photometric selection criteria that greatly optimizes the identification of quasar lenses for future searches. 
% is proposed (Eq. \ref{eq:SelectionCondition}), which separates well between the population of known lenses, the lens candidates, and the discarded lens candidates. This is found exploratively, but seems physically motivated.

We studied in detail one exemplary quasar lens system found in this survey, DR3Gaia2107-1611, based on spectroscopic observations over 3 epochs ($\sim 18$ months). We determine a lens model of the foreground source to determine the time delay between the various mirror image sightlines. We use this to convert the observed time frame to the intrinsic quasar time, covering a nearly monthly cadence. The time series of emission line characteristics show a temporal variation in two sightlines, which is detached from the third sightline. This detachment is interpreted as potentially originating from microlensing by small local variations in the density field of the foreground galaxy. Finally, we theorized that the spatial and temporal variation seen in the spectra of the quasar lenses images must be due to an intrinsic change in the emission properties of the quasar. 
%The temporal variation is believed to be due to either an intrinsic quasar change or what part of the quasar dominates the sight line.

\begin{acknowledgements}
%We would like to thank Johan Richard and Jens-Kristian Krogager for comments. 
This work has received funding from the Swiss State Secretariat for Education, Research and Innovation (SERI) under contract number MB22.00072.
The Cosmic Dawn Center (DAWN) is funded by the Danish National Research Foundation under grant DNRF140.

\end{acknowledgements}

\bibliography{references}

@ARTICLE{2023_Dux,
       author = {{Dux}, Fr{\'e}d{\'e}ric and {Lemon}, Cameron and {Courbin}, Fr{\'e}d{\'e}ric and {Sluse}, Dominique and {Smette}, Alain and {Anguita}, Timo and {Neira}, Favio},
        title = "{PS J2107{\ensuremath{-}}1611: A new wide-separation, quadruply imaged lensed quasar with flux ratio anomalies}",
      journal = {\aap},
         year = 2023,
        month = nov,
       volume = {679},
          eid = {L4},
        pages = {L4},
          doi = {10.1051/0004-6361/202348227},
archivePrefix = {arXiv},
       eprint = {2310.04494},
 primaryClass = {astro-ph.CO},
       adsurl = {https://ui.adsabs.harvard.edu/abs/2023A&A...679L...4D}
}

@article{2004_H0_from_lens,
    author = {Tortora, C. and Piedipalumbo, E. and Cardone, V. F.},
    title = {Lens modelling and estimate of H0 in quadruply lensed systems},
    journal = {Monthly Notices of the Royal Astronomical Society},
    volume = {354},
    number = {2},
    pages = {343-354},
    year = {2004},
    month = {10},
    issn = {0035-8711},
    doi = {10.1111/j.1365-2966.2004.08160.x},
    url = {https://doi.org/10.1111/j.1365-2966.2004.08160.x},
    eprint = {https://academic.oup.com/mnras/article-pdf/354/2/343/3863168/354-2-343.pdf},
}

@article{ 2015_H0_from_lens,
	author = {{Rathna Kumar, S.} and {Stalin, C. S.} and {Prabhu, T. P.}},
	title = {H0 from ten well-measured time delay lenses},
	DOI= "10.1051/0004-6361/201423977",
	url= "https://doi.org/10.1051/0004-6361/201423977",
	journal = {A\&A},
	year = 2015,
	volume = 580,
	pages = "A38",
	month = "",
}

@INPROCEEDINGS{2020_H0_from_lens,
       author = {{Wong}, Kenneth C. and {Suyu}, Sherry H. and {Chen}, Geoff C.-F. and {Rusu}, Cristian E. and {Millon}, Martin and {Sluse}, Dominique and {Bonvin}, Vivien and {Fassnacht}, Christopher D. and {Taubenberger}, Stefan and {Auger}, Matthew W. and {Birrer}, Simon and {Chan}, James H.~H. and {Courbin}, Frederic and {Hilbert}, Stefan and {Tihhonova}, Olga and {Treu}, Tommaso and {Agnello}, Adriano and {Ding}, Xuheng and {Jee}, Inh and {Komatsu}, Eiichiro and {Shajib}, Anowar J. and {Sonnenfeld}, Alessandro and {Blandford}, Roger D. and {Koopmans}, Leon V.~E. and {Marshall}, Philip J. and {Meylan}, Georges},
        title = "{H\_0 from Lensed Quasars}",
    booktitle = {H$_{0}$2020: Assessing Uncertainties in Hubble's Constant Across the Universe},
         year = 2020,
        month = oct,
          eid = {27},
        pages = {27},
          doi = {10.5281/zenodo.4062127},
       adsurl = {https://ui.adsabs.harvard.edu/abs/2020auhc.confE..27W}
}

@ARTICLE{PanSTARRS,
       author = {{Chambers}, K.~C. and {Magnier}, E.~A. and {Metcalfe}, N. and {Flewelling}, H.~A. and {Huber}, M.~E. and {Waters}, C.~Z. and {Denneau}, L. and {Draper}, P.~W. and {Farrow}, D. and {Finkbeiner}, D.~P. and {Holmberg}, C. and {Koppenhoefer}, J. and {Price}, P.~A. and {Rest}, A. and {Saglia}, R.~P. and {Schlafly}, E.~F. and {Smartt}, S.~J. and {Sweeney}, W. and {Wainscoat}, R.~J. and {Burgett}, W.~S. and {Chastel}, S. and {Grav}, T. and {Heasley}, J.~N. and {Hodapp}, K.~W. and {Jedicke}, R. and {Kaiser}, N. and {Kudritzki}, R.-P. and {Luppino}, G.~A. and {Lupton}, R.~H. and {Monet}, D.~G. and {Morgan}, J.~S. and {Onaka}, P.~M. and {Shiao}, B. and {Stubbs}, C.~W. and {Tonry}, J.~L. and {White}, R. and {Ba{\~n}ados}, E. and {Bell}, E.~F. and {Bender}, R. and {Bernard}, E.~J. and {Boegner}, M. and {Boffi}, F. and {Botticella}, M.~T. and {Calamida}, A. and {Casertano}, S. and {Chen}, W.-P. and {Chen}, X. and {Cole}, S. and {Deacon}, N. and {Frenk}, C. and {Fitzsimmons}, A. and {Gezari}, S. and {Gibbs}, V. and {Goessl}, C. and {Goggia}, T. and {Gourgue}, R. and {Goldman}, B. and {Grant}, P. and {Grebel}, E.~K. and {Hambly}, N.~C. and {Hasinger}, G. and {Heavens}, A.~F. and {Heckman}, T.~M. and {Henderson}, R. and {Henning}, T. and {Holman}, M. and {Hopp}, U. and {Ip}, W.-H. and {Isani}, S. and {Jackson}, M. and {Keyes}, C.~D. and {Koekemoer}, A.~M. and {Kotak}, R. and {Le}, D. and {Liska}, D. and {Long}, K.~S. and {Lucey}, J.~R. and {Liu}, M. and {Martin}, N.~F. and {Masci}, G. and {McLean}, B. and {Mindel}, E. and {Misra}, P. and {Morganson}, E. and {Murphy}, D.~N.~A. and {Obaika}, A. and {Narayan}, G. and {Nieto-Santisteban}, M.~A. and {Norberg}, P. and {Peacock}, J.~A. and {Pier}, E.~A. and {Postman}, M. and {Primak}, N. and {Rae}, C. and {Rai}, A. and {Riess}, A. and {Riffeser}, A. and {Rix}, H.~W. and {R{\"o}ser}, S. and {Russel}, R. and {Rutz}, L. and {Schilbach}, E. and {Schultz}, A.~S.~B. and {Scolnic}, D. and {Strolger}, L. and {Szalay}, A. and {Seitz}, S. and {Small}, E. and {Smith}, K.~W. and {Soderblom}, D.~R. and {Taylor}, P. and {Thomson}, R. and {Taylor}, A.~N. and {Thakar}, A.~R. and {Thiel}, J. and {Thilker}, D. and {Unger}, D. and {Urata}, Y. and {Valenti}, J. and {Wagner}, J. and {Walder}, T. and {Walter}, F. and {Watters}, S.~P. and {Werner}, S. and {Wood-Vasey}, W.~M. and {Wyse}, R.},
        title = "{The Pan-STARRS1 Surveys}",
      journal = {arXiv e-prints},
         year = 2016,
        month = dec,
          eid = {arXiv:1612.05560},
        pages = {arXiv:1612.05560},
          doi = {10.48550/arXiv.1612.05560},
archivePrefix = {arXiv},
       eprint = {1612.05560},
 primaryClass = {astro-ph.IM},
       adsurl = {https://ui.adsabs.harvard.edu/abs/2016arXiv161205560C}
}

@ARTICLE{Selsing16,
       author = {{Selsing}, J. and {Fynbo}, J.~P.~U. and {Christensen}, L. and {Krogager}, J.-K.},
        title = "{An X-Shooter composite of bright 1 < z < 2 quasars from UV to infrared}",
      journal = {\aap},
         year = 2016,
        month = jan,
       volume = {585},
          eid = {A87},
        pages = {A87},
          doi = {10.1051/0004-6361/201527096},
archivePrefix = {arXiv},
       eprint = {1510.08058},
 primaryClass = {astro-ph.GA},
       adsurl = {https://ui.adsabs.harvard.edu/abs/2016A\&A...585A..87S}
}

@article{MacLeod_2016,
   title={A systematic search for changing-look quasars in SDSS},
   volume={457},
   ISSN={1365-2966},
   url={http://dx.doi.org/10.1093/mnras/stv2997},
   DOI={10.1093/mnras/stv2997},
   number={1},
   journal={Monthly Notices of the Royal Astronomical Society},
   publisher={Oxford University Press (OUP)},
   author={MacLeod, Chelsea L. and Ross, Nicholas P. and Lawrence, Andy and Goad, Mike and Horne, Keith and Burgett, William and Chambers, Ken C. and Flewelling, Heather and Hodapp, Klaus and Kaiser, Nick and Magnier, Eugene and Wainscoat, Richard and Waters, Christopher},
   year={2016},
   month=jan, pages={389–404} }

@misc{2015_Runnoe,
      title={Now You See It, Now You Don't: The Disappearing Central Engine of the Quasar J1011+5442}, 
      author={Jessie C. Runnoe and Sabrina Cales and John J. Ruan and Michael Eracleous and Scott F. Anderson and Yue Shen and Paul Green and Eric Morganson and Stephanie LaMassa and Jenny E. Greene and Tom Dwelly and Donald P. Schneider and Andrea Merloni and Antonis Georgakakis},
      year={2015},
      eprint={1509.03640},
      archivePrefix={arXiv},
      primaryClass={astro-ph.GA},
      url={https://arxiv.org/abs/1509.03640}, 
}

@ARTICLE{2015_LaMassa,
       author = {{LaMassa}, Stephanie M. and {Cales}, Sabrina and {Moran}, Edward C. and {Myers}, Adam D. and {Richards}, Gordon T. and {Eracleous}, Michael and {Heckman}, Timothy M. and {Gallo}, Luigi and {Urry}, C. Megan},
        title = "{The Discovery of the First {\textquotedblleft}Changing Look{\textquotedblright} Quasar: New Insights Into the Physics and Phenomenology of Active Galactic Nucleus}",
      journal = {\apj},
         year = 2015,
        month = feb,
       volume = {800},
       number = {2},
          eid = {144},
        pages = {144},
          doi = {10.1088/0004-637X/800/2/144},
archivePrefix = {arXiv},
       eprint = {1412.2136},
 primaryClass = {astro-ph.GA},
       adsurl = {https://ui.adsabs.harvard.edu/abs/2015ApJ...800..144L}
}

@article{Chae_2013,
   title={Modelling mass distribution in elliptical galaxies: mass profiles and their correlation with velocity dispersion profiles},
   volume={437},
   ISSN={1365-2966},
   url={http://dx.doi.org/10.1093/mnras/stt2163},
   DOI={10.1093/mnras/stt2163},
   number={4},
   journal={Monthly Notices of the Royal Astronomical Society},
   publisher={Oxford University Press (OUP)},
   author={Chae, Kyu-Hyun and Bernardi, Mariangela and Kravtsov, Andrey V.},
   year={2013},
   month=dec, pages={3670–3687} }

@article{Oguri_2010,
   title={Gravitationally lensed quasars and supernovae in future wide-field optical imaging surveys: Lensed quasars and supernovae},
   ISSN={1365-2966},
   url={http://dx.doi.org/10.1111/j.1365-2966.2010.16639.x},
   DOI={10.1111/j.1365-2966.2010.16639.x},
   journal={Monthly Notices of the Royal Astronomical Society},
   publisher={Oxford University Press (OUP)},
   author={Oguri, Masamune and Marshall, Philip J.},
   year={2010},
   month=apr, pages={no-no} }

@article{Eulaers_2011,
	author = {{Eulaers, E.} and {Magain, P.}},
	title = {Time delays for eleven gravitationally lensed quasars revisited},
	DOI= "10.1051/0004-6361/201016101",
	url= "https://doi.org/10.1051/0004-6361/201016101",
	journal = {A\&A},
	year = 2011,
	volume = 536,
	pages = "A44",
	month = "",
}

@article{Coatman_2016,
   title={Civ emission-line properties and systematic trends in quasar black hole mass estimates},
   volume={461},
   ISSN={1365-2966},
   url={http://dx.doi.org/10.1093/mnras/stw1360},
   DOI={10.1093/mnras/stw1360},
   number={1},
   journal={Monthly Notices of the Royal Astronomical Society},
   publisher={Oxford University Press (OUP)},
   author={Coatman, Liam and Hewett, Paul C. and Banerji, Manda and Richards, Gordon T.},
   year={2016},
   month=jun, pages={647–665} }

@article{Wang_2012,
   title={METALLICITY AND QUASAR OUTFLOWS},
   volume={751},
   ISSN={2041-8213},
   url={http://dx.doi.org/10.1088/2041-8205/751/2/L23},
   DOI={10.1088/2041-8205/751/2/l23},
   number={2},
   journal={The Astrophysical Journal},
   publisher={American Astronomical Society},
   author={Wang, Huiyuan and Zhou, Hongyan and Yuan, Weimin and Wang, Tinggui},
   year={2012},
   month=may, pages={L23} }

@article{Temple_2021,
   title={High-ionization emission-line ratios from quasar broad-line regions: metallicity or density?},
   volume={505},
   ISSN={1365-2966},
   url={http://dx.doi.org/10.1093/mnras/stab1610},
   DOI={10.1093/mnras/stab1610},
   number={3},
   journal={Monthly Notices of the Royal Astronomical Society},
   publisher={Oxford University Press (OUP)},
   author={Temple, Matthew J and Ferland, Gary J and Rankine, Amy L and Chatzikos, Marios and Hewett, Paul C},
   year={2021},
   month=jun, pages={3247–3259} }

@INPROCEEDINGS{2012_Richards,
       author = {{Richards}, G.~T.},
        title = "{CIV Emission as a Probe of Accretion Disk Winds}",
    booktitle = {AGN Winds in Charleston},
         year = 2012,
       editor = {{Chartas}, G. and {Hamann}, F. and {Leighly}, K.~M.},
       series = {Astronomical Society of the Pacific Conference Series},
       volume = {460},
        month = aug,
        pages = {67},
          doi = {10.48550/arXiv.1201.2595},
archivePrefix = {arXiv},
       eprint = {1201.2595},
 primaryClass = {astro-ph.CO},
       adsurl = {https://ui.adsabs.harvard.edu/abs/2012ASPC..460...67R}
}

@ARTICLE{2019_Krogager,
       author = {{Krogager}, Jens-Kristian and {Fynbo}, Johan P.~U. and {M{\o}ller}, Palle and {Noterdaeme}, Pasquier and {Heintz}, Kasper E. and {Pettini}, Max},
        title = "{The effect of dust bias on the census of neutral gas and metals in the high-redshift Universe due to SDSS-II quasar colour selection}",
      journal = {\mnras},
         year = 2019,
        month = jul,
       volume = {486},
       number = {3},
        pages = {4377-4397},
          doi = {10.1093/mnras/stz1120},
archivePrefix = {arXiv},
       eprint = {1904.06966},
 primaryClass = {astro-ph.GA},
       adsurl = {https://ui.adsabs.harvard.edu/abs/2019MNRAS.486.4377K}
}

@ARTICLE{2020_Planck,
       author = {{Planck Collaboration} and {Aghanim}, N. and {Akrami}, Y. and {Ashdown}, M. and {Aumont}, J. and {Baccigalupi}, C. and {Ballardini}, M. and {Banday}, A.~J. and {Barreiro}, R.~B. and {Bartolo}, N. and {Basak}, S. and {Battye}, R. and {Benabed}, K. and {Bernard}, J. -P. and {Bersanelli}, M. and {Bielewicz}, P. and {Bock}, J.~J. and {Bond}, J.~R. and {Borrill}, J. and {Bouchet}, F.~R. and {Boulanger}, F. and {Bucher}, M. and {Burigana}, C. and {Butler}, R.~C. and {Calabrese}, E. and {Cardoso}, J. -F. and {Carron}, J. and {Challinor}, A. and {Chiang}, H.~C. and {Chluba}, J. and {Colombo}, L.~P.~L. and {Combet}, C. and {Contreras}, D. and {Crill}, B.~P. and {Cuttaia}, F. and {de Bernardis}, P. and {de Zotti}, G. and {Delabrouille}, J. and {Delouis}, J. -M. and {Di Valentino}, E. and {Diego}, J.~M. and {Dor{\'e}}, O. and {Douspis}, M. and {Ducout}, A. and {Dupac}, X. and {Dusini}, S. and {Efstathiou}, G. and {Elsner}, F. and {En{\ss}lin}, T.~A. and {Eriksen}, H.~K. and {Fantaye}, Y. and {Farhang}, M. and {Fergusson}, J. and {Fernandez-Cobos}, R. and {Finelli}, F. and {Forastieri}, F. and {Frailis}, M. and {Fraisse}, A.~A. and {Franceschi}, E. and {Frolov}, A. and {Galeotta}, S. and {Galli}, S. and {Ganga}, K. and {G{\'e}nova-Santos}, R.~T. and {Gerbino}, M. and {Ghosh}, T. and {Gonz{\'a}lez-Nuevo}, J. and {G{\'o}rski}, K.~M. and {Gratton}, S. and {Gruppuso}, A. and {Gudmundsson}, J.~E. and {Hamann}, J. and {Handley}, W. and {Hansen}, F.~K. and {Herranz}, D. and {Hildebrandt}, S.~R. and {Hivon}, E. and {Huang}, Z. and {Jaffe}, A.~H. and {Jones}, W.~C. and {Karakci}, A. and {Keih{\"a}nen}, E. and {Keskitalo}, R. and {Kiiveri}, K. and {Kim}, J. and {Kisner}, T.~S. and {Knox}, L. and {Krachmalnicoff}, N. and {Kunz}, M. and {Kurki-Suonio}, H. and {Lagache}, G. and {Lamarre}, J. -M. and {Lasenby}, A. and {Lattanzi}, M. and {Lawrence}, C.~R. and {Le Jeune}, M. and {Lemos}, P. and {Lesgourgues}, J. and {Levrier}, F. and {Lewis}, A. and {Liguori}, M. and {Lilje}, P.~B. and {Lilley}, M. and {Lindholm}, V. and {L{\'o}pez-Caniego}, M. and {Lubin}, P.~M. and {Ma}, Y. -Z. and {Mac{\'\i}as-P{\'e}rez}, J.~F. and {Maggio}, G. and {Maino}, D. and {Mandolesi}, N. and {Mangilli}, A. and {Marcos-Caballero}, A. and {Maris}, M. and {Martin}, P.~G. and {Martinelli}, M. and {Mart{\'\i}nez-Gonz{\'a}lez}, E. and {Matarrese}, S. and {Mauri}, N. and {McEwen}, J.~D. and {Meinhold}, P.~R. and {Melchiorri}, A. and {Mennella}, A. and {Migliaccio}, M. and {Millea}, M. and {Mitra}, S. and {Miville-Desch{\^e}nes}, M. -A. and {Molinari}, D. and {Montier}, L. and {Morgante}, G. and {Moss}, A. and {Natoli}, P. and {N{\o}rgaard-Nielsen}, H.~U. and {Pagano}, L. and {Paoletti}, D. and {Partridge}, B. and {Patanchon}, G. and {Peiris}, H.~V. and {Perrotta}, F. and {Pettorino}, V. and {Piacentini}, F. and {Polastri}, L. and {Polenta}, G. and {Puget}, J. -L. and {Rachen}, J.~P. and {Reinecke}, M. and {Remazeilles}, M. and {Renzi}, A. and {Rocha}, G. and {Rosset}, C. and {Roudier}, G. and {Rubi{\~n}o-Mart{\'\i}n}, J.~A. and {Ruiz-Granados}, B. and {Salvati}, L. and {Sandri}, M. and {Savelainen}, M. and {Scott}, D. and {Shellard}, E.~P.~S. and {Sirignano}, C. and {Sirri}, G. and {Spencer}, L.~D. and {Sunyaev}, R. and {Suur-Uski}, A. -S. and {Tauber}, J.~A. and {Tavagnacco}, D. and {Tenti}, M. and {Toffolatti}, L. and {Tomasi}, M. and {Trombetti}, T. and {Valenziano}, L. and {Valiviita}, J. and {Van Tent}, B. and {Vibert}, L. and {Vielva}, P. and {Villa}, F. and {Vittorio}, N. and {Wandelt}, B.~D. and {Wehus}, I.~K. and {White}, M. and {White}, S.~D.~M. and {Zacchei}, A. and {Zonca}, A.},
        title = "{Planck 2018 results. VI. Cosmological parameters}",
      journal = {\aap},
         year = 2020,
        month = sep,
       volume = {641},
          eid = {A6},
        pages = {A6},
          doi = {10.1051/0004-6361/201833910},
archivePrefix = {arXiv},
       eprint = {1807.06209},
 primaryClass = {astro-ph.CO},
       adsurl = {https://ui.adsabs.harvard.edu/abs/2020A\&A...641A...6P}
}

@ARTICLE{2018_Krogager,
       author = {{Krogager}, J. -K. and {Noterdaeme}, P. and {O'Meara}, J.~M. and {Fumagalli}, M. and {Fynbo}, J.~P.~U. and {Prochaska}, J.~X. and {Hennawi}, J. and {Balashev}, S. and {Courbin}, F. and {Rafelski}, M. and {Smette}, A. and {Boiss{\'e}}, P.},
        title = "{Dissecting cold gas in a high-redshift galaxy using a lensed background quasar}",
      journal = {\aap},
         year = 2018,
        month = nov,
       volume = {619},
          eid = {A142},
        pages = {A142},
          doi = {10.1051/0004-6361/201833608},
archivePrefix = {arXiv},
       eprint = {1809.01053},
 primaryClass = {astro-ph.GA},
       adsurl = {https://ui.adsabs.harvard.edu/abs/2018A\&A...619A.142K}
}

@article{Nightingale_2021,
   title={PyAutoLens: Open-Source Strong Gravitational Lensing},
   volume={6},
   ISSN={2475-9066},
   url={http://dx.doi.org/10.21105/joss.02825},
   DOI={10.21105/joss.02825},
   number={58},
   journal={Journal of Open Source Software},
   publisher={The Open Journal},
   author={Nightingale, James. and Hayes, Richard and Kelly, Ashley and Amvrosiadis, Aristeidis and Etherington, Amy and He, Qiuhan and Li, Nan and Cao, XiaoYue and Frawley, Jonathan and Cole, Shaun and Enia, Andrea and Frenk, Carlos and Harvey, David and Li, Ran and Massey, Richard and Negrello, Mattia and Robertson, Andrew},
   year={2021},
   month=feb, pages={2825} }

@ARTICLE{2012_Volonteri,
       author = {{Volonteri}, M.},
        title = "{The Formation and Evolution of Massive Black Holes}",
      journal = {Science},
         year = 2012,
        month = aug,
       volume = {337},
       number = {6094},
        pages = {544},
          doi = {10.1126/science.1220843},
archivePrefix = {arXiv},
       eprint = {1208.1106},
 primaryClass = {astro-ph.CO},
       adsurl = {https://ui.adsabs.harvard.edu/abs/2012Sci...337..544V}
}

@misc{gtc_osiris,
  author       = {{Gran Telescopio Canarias}},
  title        = {{OSIRIS Instrument Overview}},
  howpublished = {\url{https://www.gtc.iac.es/instruments/osiris/}},
  note         = {Accessed: 2025-06-11},
  year         = {2025}
}

@article{2014_AllWise,
       author = {{Cutri}, R.~M. and {Wright}, E.~L. and {Conrow}, T. and {Fowler}, J.~W. and {Eisenhardt}, P.~R.~M. and {Grillmair}, C. and {Kirkpatrick}, J.~D. and {Masci}, F. and {McCallon}, H.~L. and {Wheelock}, S.~L. and {Fajardo-Acosta}, S. and {Yan}, L. and {Benford}, D. and {Harbut}, M. and {Jarrett}, T. and {Lake}, S. and {Leisawitz}, D. and {Ressler}, M.~E. and {Stanford}, S.~A. and {Tsai}, C. -W. and {Liu}, F. and {Helou}, G. and {Mainzer}, A. and {Gettngs}, D. and {Gonzalez}, A. and {Hoffman}, D. and {Marsh}, K.~A. and {Padgett}, D. and {Skrutskie}, M.~F. and {Beck}, R. and {Papin}, M. and {Wittman}, M.},
        title = "{VizieR Online Data Catalog: AllWISE Data Release (Cutri+ 2013)}",
 howpublished = {VizieR On-line Data Catalog: II/328.  Originally published in: IPAC/Caltech (2013)},
         year = 2021,
        month = feb,
          eid = {II/328},
       adsurl = {https://ui.adsabs.harvard.edu/abs/2014yCat.2328....0C},
    journal= "IPAC/Caltech"
}

@ARTICLE{2006_2MASS,
       author = {{Skrutskie}, M.~F. and {Cutri}, R.~M. and {Stiening}, R. and {Weinberg}, M.~D. and {Schneider}, S. and {Carpenter}, J.~M. and {Beichman}, C. and {Capps}, R. and {Chester}, T. and {Elias}, J. and {Huchra}, J. and {Liebert}, J. and {Lonsdale}, C. and {Monet}, D.~G. and {Price}, S. and {Seitzer}, P. and {Jarrett}, T. and {Kirkpatrick}, J.~D. and {Gizis}, J.~E. and {Howard}, E. and {Evans}, T. and {Fowler}, J. and {Fullmer}, L. and {Hurt}, R. and {Light}, R. and {Kopan}, E.~L. and {Marsh}, K.~A. and {McCallon}, H.~L. and {Tam}, R. and {Van Dyk}, S. and {Wheelock}, S.},
        title = "{The Two Micron All Sky Survey (2MASS)}",
      journal = {\aj},
         year = 2006,
        month = feb,
       volume = {131},
       number = {2},
        pages = {1163-1183},
          doi = {10.1086/498708},
       adsurl = {https://ui.adsabs.harvard.edu/abs/2006AJ....131.1163S}
}

@misc{chambers2019panstarrs1surveys,
      title={The Pan-STARRS1 Surveys}, 
      author={K. C. Chambers and E. A. Magnier and N. Metcalfe and H. A. Flewelling and M. E. Huber and C. Z. Waters and L. Denneau and P. W. Draper and D. Farrow and D. P. Finkbeiner and C. Holmberg and J. Koppenhoefer and P. A. Price and A. Rest and R. P. Saglia and E. F. Schlafly and S. J. Smartt and W. Sweeney and R. J. Wainscoat and W. S. Burgett and S. Chastel and T. Grav and J. N. Heasley and K. W. Hodapp and R. Jedicke and N. Kaiser and R. -P. Kudritzki and G. A. Luppino and R. H. Lupton and D. G. Monet and J. S. Morgan and P. M. Onaka and B. Shiao and C. W. Stubbs and J. L. Tonry and R. White and E. Bañados and E. F. Bell and R. Bender and E. J. Bernard and M. Boegner and F. Boffi and M. T. Botticella and A. Calamida and S. Casertano and W. -P. Chen and X. Chen and S. Cole and N. Deacon and C. Frenk and A. Fitzsimmons and S. Gezari and V. Gibbs and C. Goessl and T. Goggia and R. Gourgue and B. Goldman and P. Grant and E. K. Grebel and N. C. Hambly and G. Hasinger and A. F. Heavens and T. M. Heckman and R. Henderson and T. Henning and M. Holman and U. Hopp and W. -H. Ip and S. Isani and M. Jackson and C. D. Keyes and A. M. Koekemoer and R. Kotak and D. Le and D. Liska and K. S. Long and J. R. Lucey and M. Liu and N. F. Martin and G. Masci and B. McLean and E. Mindel and P. Misra and E. Morganson and D. N. A. Murphy and A. Obaika and G. Narayan and M. A. Nieto-Santisteban and P. Norberg and J. A. Peacock and E. A. Pier and M. Postman and N. Primak and C. Rae and A. Rai and A. Riess and A. Riffeser and H. W. Rix and S. Röser and R. Russel and L. Rutz and E. Schilbach and A. S. B. Schultz and D. Scolnic and L. Strolger and A. Szalay and S. Seitz and E. Small and K. W. Smith and D. R. Soderblom and P. Taylor and R. Thomson and A. N. Taylor and A. R. Thakar and J. Thiel and D. Thilker and D. Unger and Y. Urata and J. Valenti and J. Wagner and T. Walder and F. Walter and S. P. Watters and S. Werner and W. M. Wood-Vasey and R. Wyse},
      year={2019},
      eprint={1612.05560},
      archivePrefix={arXiv},
      primaryClass={astro-ph.IM},
      url={https://arxiv.org/abs/1612.05560}, 
}

@ARTICLE{2002_SDSS,
       author = {{Stoughton}, Chris and {Lupton}, Robert H. and {Bernardi}, Mariangela and {Blanton}, Michael R. and {Burles}, Scott and {Castander}, Francisco J. and {Connolly}, A.~J. and {Eisenstein}, Daniel J. and {Frieman}, Joshua A. and {Hennessy}, G.~S. and {Hindsley}, Robert B. and {Ivezi{\'c}}, {\v{Z}}eljko and {Kent}, Stephen and {Kunszt}, Peter Z. and {Lee}, Brian C. and {Meiksin}, Avery and {Munn}, Jeffrey A. and {Newberg}, Heidi Jo and {Nichol}, R.~C. and {Nicinski}, Tom and {Pier}, Jeffrey R. and {Richards}, Gordon T. and {Richmond}, Michael W. and {Schlegel}, David J. and {Smith}, J. Allyn and {Strauss}, Michael A. and {SubbaRao}, Mark and {Szalay}, Alexander S. and {Thakar}, Aniruddha R. and {Tucker}, Douglas L. and {Vanden Berk}, Daniel E. and {Yanny}, Brian and {Adelman}, Jennifer K. and {Anderson}, Jr., John E. and {Anderson}, Scott F. and {Annis}, James and {Bahcall}, Neta A. and {Bakken}, J.~A. and {Bartelmann}, Matthias and {Bastian}, Steven and {Bauer}, Amanda and {Berman}, Eileen and {B{\"o}hringer}, Hans and {Boroski}, William N. and {Bracker}, Steve and {Briegel}, Charlie and {Briggs}, John W. and {Brinkmann}, J. and {Brunner}, Robert and {Carey}, Larry and {Carr}, Michael A. and {Chen}, Bing and {Christian}, Damian and {Colestock}, Patrick L. and {Crocker}, J.~H. and {Csabai}, Istv{\'a}n and {Czarapata}, Paul C. and {Dalcanton}, Julianne and {Davidsen}, Arthur F. and {Davis}, John Eric and {Dehnen}, Walter and {Dodelson}, Scott and {Doi}, Mamoru and {Dombeck}, Tom and {Donahue}, Megan and {Ellman}, Nancy and {Elms}, Brian R. and {Evans}, Michael L. and {Eyer}, Laurent and {Fan}, Xiaohui and {Federwitz}, Glenn R. and {Friedman}, Scott and {Fukugita}, Masataka and {Gal}, Roy and {Gillespie}, Bruce and {Glazebrook}, Karl and {Gray}, Jim and {Grebel}, Eva K. and {Greenawalt}, Bruce and {Greene}, Gretchen and {Gunn}, James E. and {de Haas}, Ernst and {Haiman}, Zolt{\'a}n and {Haldeman}, Merle and {Hall}, Patrick B. and {Hamabe}, Masaru and {Hansen}, Brad and {Harris}, Frederick H. and {Harris}, Hugh and {Harvanek}, Michael and {Hawley}, Suzanne L. and {Hayes}, J.~J.~E. and {Heckman}, Timothy M. and {Helmi}, Amina and {Henden}, Arne and {Hogan}, Craig J. and {Hogg}, David W. and {Holmgren}, Donald J. and {Holtzman}, Jon and {Huang}, Chih-Hao and {Hull}, Charles and {Ichikawa}, Shin-Ichi and {Ichikawa}, Takashi and {Johnston}, David E. and {Kauffmann}, Guinevere and {Kim}, Rita S.~J. and {Kimball}, Tim and {Kinney}, E. and {Klaene}, Mark and {Kleinman}, S.~J. and {Klypin}, Anatoly and {Knapp}, G.~R. and {Korienek}, John and {Krolik}, Julian and {Kron}, Richard G. and {Krzesi{\'n}ski}, Jurek and {Lamb}, D.~Q. and {Leger}, R. French and {Limmongkol}, Siriluk and {Lindenmeyer}, Carl and {Long}, Daniel C. and {Loomis}, Craig and {Loveday}, Jon and {MacKinnon}, Bryan and {Mannery}, Edward J. and {Mantsch}, P.~M. and {Margon}, Bruce and {McGehee}, Peregrine and {McKay}, Timothy A. and {McLean}, Brian and {Menou}, Kristen and {Merelli}, Aronne and {Mo}, H.~J. and {Monet}, David G. and {Nakamura}, Osamu and {Narayanan}, Vijay K. and {Nash}, Thomas and {Neilsen}, Jr., Eric H. and {Newman}, Peter R. and {Nitta}, Atsuko and {Odenkirchen}, Michael and {Okada}, Norio and {Okamura}, Sadanori and {Ostriker}, Jeremiah P. and {Owen}, Russell and {Pauls}, A. George and {Peoples}, John and {Peterson}, R.~S. and {Petravick}, Donald and {Pope}, Adrian and {Pordes}, Ruth and {Postman}, Marc and {Prosapio}, Angela and {Quinn}, Thomas R. and {Rechenmacher}, Ron and {Rivetta}, Claudio H. and {Rix}, Hans-Walter and {Rockosi}, Constance M. and {Rosner}, Robert and {Ruthmansdorfer}, Kurt and {Sandford}, Dale and {Schneider}, Donald P. and {Scranton}, Ryan and {Sekiguchi}, Maki and {Sergey}, Gary and {Sheth}, Ravi and {Shimasaku}, Kazuhiro and {Smee}, Stephen and {Snedden}, Stephanie A. and {Stebbins}, Albert and {Stubbs}, Christopher and {Szapudi}, Istv{\'a}n and {Szkody}, Paula and {Szokoly}, Gyula P. and {Tabachnik}, Serge and {Tsvetanov}, Zlatan and {Uomoto}, Alan and {Vogeley}, Michael S. and {Voges}, Wolfgang and {Waddell}, Patrick and {Walterbos}, Ren{\'e} and {Wang}, Shu-i. and {Watanabe}, Masaru and {Weinberg}, David H. and {White}, Richard L. and {White}, Simon D.~M. and {Wilhite}, Brian and {Wolfe}, David and {Yasuda}, Naoki and {York}, Donald G. and {Zehavi}, Idit and {Zheng}, Wei},
    title = "{Sloan Digital Sky Survey: Early Data Release}",
    journal = {\aj},
         year = 2002,
        month = jan,
       volume = {123},
       number = {1},
        pages = {485-548},
          doi = {10.1086/324741},
       adsurl = {https://ui.adsabs.harvard.edu/abs/2002AJ....123..485S}
}

@ARTICLE{2004_Richards,
       author = {{Richards}, Gordon T. and {Nichol}, Robert C. and {Gray}, Alexander G. and {Brunner}, Robert J. and {Lupton}, Robert H. and {Vanden Berk}, Daniel E. and {Chong}, Shang Shan and {Weinstein}, Michael A. and {Schneider}, Donald P. and {Anderson}, Scott F. and {Munn}, Jeffrey A. and {Harris}, Hugh C. and {Strauss}, Michael A. and {Fan}, Xiaohui and {Gunn}, James E. and {Ivezi{\'c}}, {\v{Z}}eljko and {York}, Donald G. and {Brinkmann}, J. and {Moore}, Andrew W.},
        title = "{Efficient Photometric Selection of Quasars from the Sloan Digital Sky Survey: 100,000 z < 3 Quasars from Data Release One}",
      journal = {\apjs},
         year = 2004,
        month = dec,
       volume = {155},
       number = {2},
        pages = {257-269},
          doi = {10.1086/425356},
archivePrefix = {arXiv},
       eprint = {astro-ph/0408505},
 primaryClass = {astro-ph},
       adsurl = {https://ui.adsabs.harvard.edu/abs/2004ApJS..155..257R}
}

@ARTICLE{2015_Krawczyk,
       author = {{Krawczyk}, Coleman M. and {Richards}, Gordon T. and {Gallagher}, S.~C. and {Leighly}, Karen M. and {Hewett}, Paul C. and {Ross}, Nicholas P. and {Hall}, P.~B.},
        title = "{Mining for Dust in Type 1 Quasars}",
      journal = {\aj},
         year = 2015,
        month = jun,
       volume = {149},
       number = {6},
          eid = {203},
        pages = {203},
          doi = {10.1088/0004-6256/149/6/203},
archivePrefix = {arXiv},
       eprint = {1412.7039},
 primaryClass = {astro-ph.GA},
       adsurl = {https://ui.adsabs.harvard.edu/abs/2015AJ....149..203K}
}

@ARTICLE{2018_GaiaDR2,
       author = {{Lindegren}, L. and {Hern{\'a}ndez}, J. and {Bombrun}, A. and {Klioner}, S. and {Bastian}, U. and {Ramos-Lerate}, M. and {de Torres}, A. and {Steidelm{\"u}ller}, H. and {Stephenson}, C. and {Hobbs}, D. and {Lammers}, U. and {Biermann}, M. and {Geyer}, R. and {Hilger}, T. and {Michalik}, D. and {Stampa}, U. and {McMillan}, P.~J. and {Casta{\~n}eda}, J. and {Clotet}, M. and {Comoretto}, G. and {Davidson}, M. and {Fabricius}, C. and {Gracia}, G. and {Hambly}, N.~C. and {Hutton}, A. and {Mora}, A. and {Portell}, J. and {van Leeuwen}, F. and {Abbas}, U. and {Abreu}, A. and {Altmann}, M. and {Andrei}, A. and {Anglada}, E. and {Balaguer-N{\'u}{\~n}ez}, L. and {Barache}, C. and {Becciani}, U. and {Bertone}, S. and {Bianchi}, L. and {Bouquillon}, S. and {Bourda}, G. and {Br{\"u}semeister}, T. and {Bucciarelli}, B. and {Busonero}, D. and {Buzzi}, R. and {Cancelliere}, R. and {Carlucci}, T. and {Charlot}, P. and {Cheek}, N. and {Crosta}, M. and {Crowley}, C. and {de Bruijne}, J. and {de Felice}, F. and {Drimmel}, R. and {Esquej}, P. and {Fienga}, A. and {Fraile}, E. and {Gai}, M. and {Garralda}, N. and {Gonz{\'a}lez-Vidal}, J.~J. and {Guerra}, R. and {Hauser}, M. and {Hofmann}, W. and {Holl}, B. and {Jordan}, S. and {Lattanzi}, M.~G. and {Lenhardt}, H. and {Liao}, S. and {Licata}, E. and {Lister}, T. and {L{\"o}ffler}, W. and {Marchant}, J. and {Martin-Fleitas}, J. -M. and {Messineo}, R. and {Mignard}, F. and {Morbidelli}, R. and {Poggio}, E. and {Riva}, A. and {Rowell}, N. and {Salguero}, E. and {Sarasso}, M. and {Sciacca}, E. and {Siddiqui}, H. and {Smart}, R.~L. and {Spagna}, A. and {Steele}, I. and {Taris}, F. and {Torra}, J. and {van Elteren}, A. and {van Reeven}, W. and {Vecchiato}, A.},
        title = "{Gaia Data Release 2. The astrometric solution}",
      journal = {\aap},
         year = 2018,
        month = aug,
       volume = {616},
          eid = {A2},
        pages = {A2},
          doi = {10.1051/0004-6361/201832727},
archivePrefix = {arXiv},
       eprint = {1804.09366},
 primaryClass = {astro-ph.IM},
       adsurl = {https://ui.adsabs.harvard.edu/abs/2018A\&A...616A...2L}
}

@ARTICLE{2015_Heintz,
       author = {{Heintz}, K.~E. and {Fynbo}, J.~P.~U. and {H{\o}g}, E.},
        title = "{A study of purely astrometric selection of extragalactic point sources with Gaia}",
      journal = {\aap},
         year = 2015,
        month = jun,
       volume = {578},
          eid = {A91},
        pages = {A91},
          doi = {10.1051/0004-6361/201526038},
archivePrefix = {arXiv},
       eprint = {1503.02874},
 primaryClass = {astro-ph.HE},
       adsurl = {https://ui.adsabs.harvard.edu/abs/2015A\&A...578A..91H}
}

@ARTICLE{1986_Koo,
       author = {{Koo}, D.~C. and {Kron}, R.~G. and {Cudworth}, K.~M.},
        title = "{Quasars to B > 22.5 in selected area 57 : a catalog of multicolor photometry, variability and astrometry.}",
      journal = {\pasp},
         year = 1986,
        month = mar,
       volume = {98},
        pages = {285-306},
          doi = {10.1086/131756},
       adsurl = {https://ui.adsabs.harvard.edu/abs/1986PASP...98..285K}
}

@ARTICLE{1992_Pei,
       author = {{Pei}, Yichuan C.},
        title = "{Interstellar Dust from the Milky Way to the Magellanic Clouds}",
      journal = {\apj},
         year = 1992,
        month = aug,
       volume = {395},
        pages = {130},
          doi = {10.1086/171637},
       adsurl = {https://ui.adsabs.harvard.edu/abs/1992ApJ...395..130P}
}

@ARTICLE{2012_Stern,
       author = {{Stern}, Daniel and {Assef}, Roberto J. and {Benford}, Dominic J. and {Blain}, Andrew and {Cutri}, Roc and {Dey}, Arjun and {Eisenhardt}, Peter and {Griffith}, Roger L. and {Jarrett}, T.~H. and {Lake}, Sean and {Masci}, Frank and {Petty}, Sara and {Stanford}, S.~A. and {Tsai}, Chao-Wei and {Wright}, E.~L. and {Yan}, Lin and {Harrison}, Fiona and {Madsen}, Kristin},
        title = "{Mid-infrared Selection of Active Galactic Nuclei with the Wide-Field Infrared Survey Explorer. I. Characterizing WISE-selected Active Galactic Nuclei in COSMOS}",
      journal = {\apj},
         year = 2012,
        month = jul,
       volume = {753},
       number = {1},
          eid = {30},
        pages = {30},
          doi = {10.1088/0004-637X/753/1/30},
archivePrefix = {arXiv},
       eprint = {1205.0811},
 primaryClass = {astro-ph.CO},
       adsurl = {https://ui.adsabs.harvard.edu/abs/2012ApJ...753...30S}
}

@ARTICLE{2015_Secrest,
       author = {{Secrest}, N.~J. and {Dudik}, R.~P. and {Dorland}, B.~N. and {Zacharias}, N. and {Makarov}, V. and {Fey}, A. and {Frouard}, J. and {Finch}, C.},
        title = "{Identification of 1.4 Million Active Galactic Nuclei in the Mid-Infrared using WISE Data}",
      journal = {\apjs},
         year = 2015,
        month = nov,
       volume = {221},
       number = {1},
          eid = {12},
        pages = {12},
          doi = {10.1088/0067-0049/221/1/12},
archivePrefix = {arXiv},
       eprint = {1509.07289},
 primaryClass = {astro-ph.GA},
       adsurl = {https://ui.adsabs.harvard.edu/abs/2015ApJS..221...12S}
}

@ARTICLE{2012_Mateos,
       author = {{Mateos}, S. and {Alonso-Herrero}, A. and {Carrera}, F.~J. and {Blain}, A. and {Watson}, M.~G. and {Barcons}, X. and {Braito}, V. and {Severgnini}, P. and {Donley}, J.~L. and {Stern}, D.},
        title = "{Using the Bright Ultrahard XMM-Newton survey to define an IR selection of luminous AGN based on WISE colours}",
      journal = {\mnras},
         year = 2012,
        month = nov,
       volume = {426},
       number = {4},
        pages = {3271-3281},
          doi = {10.1111/j.1365-2966.2012.21843.x},
archivePrefix = {arXiv},
       eprint = {1208.2530},
 primaryClass = {astro-ph.CO},
       adsurl = {https://ui.adsabs.harvard.edu/abs/2012MNRAS.426.3271M}
}

@ARTICLE{2008_Maddox,
       author = {{Maddox}, Natasha and {Hewett}, Paul C. and {Warren}, S.~J. and {Croom}, S.~M.},
        title = "{Luminous K-band selected quasars from UKIDSS}",
      journal = {\mnras},
         year = 2008,
        month = may,
       volume = {386},
       number = {3},
        pages = {1605-1624},
          doi = {10.1111/j.1365-2966.2008.13138.x},
archivePrefix = {arXiv},
       eprint = {0802.3650},
 primaryClass = {astro-ph},
       adsurl = {https://ui.adsabs.harvard.edu/abs/2008MNRAS.386.1605M}
}

@ARTICLE{2000_Warren,
       author = {{Warren}, S.~J. and {Hewett}, P.~C. and {Foltz}, C.~B.},
        title = "{The KX method for producing K-band flux-limited samples of quasars}",
      journal = {\mnras},
         year = 2000,
        month = mar,
       volume = {312},
       number = {4},
        pages = {827-832},
          doi = {10.1046/j.1365-8711.2000.03206.x},
archivePrefix = {arXiv},
       eprint = {astro-ph/9911064},
 primaryClass = {astro-ph},
       adsurl = {https://ui.adsabs.harvard.edu/abs/2000MNRAS.312..827W}
}

@article{Schawinski_2012,
   title={Heavily obscured quasar host galaxies at z ∼ 2 are discs, not major mergers},
   volume={425},
   ISSN={1745-3933},
   url={http://dx.doi.org/10.1111/j.1745-3933.2012.01302.x},
   DOI={10.1111/j.1745-3933.2012.01302.x},
   number={1},
   journal={Monthly Notices of the Royal Astronomical Society: Letters},
   publisher={Oxford University Press (OUP)},
   author={Schawinski, Kevin and Simmons, Brooke D. and Urry, C. Megan and Treister, Ezequiel and Glikman, Eilat},
   year={2012},
   month=sep, pages={L61–L65} }

@misc{jing2025quasarpaircatalogcompiled,
      title={A Quasar Pair Catalog Compiled from DESI DR1}, 
      author={Liang Jing and Qihang Chen and Zhuojun Deng and Xingyu Zhu and Hu Zou and Jianghua Wu},
      year={2025},
      eprint={2505.03103},
      archivePrefix={arXiv},
      primaryClass={astro-ph.GA},
      url={https://arxiv.org/abs/2505.03103}, 
}

@misc{cambridge_lensed_quasars,
  author       = {Gravitationally Lensed Quasar Database, University of Cambridge},
  title        = {Gravitationally Lensed Quasar Database},
  year         = {2024},
  url          = {https://research.ast.cam.ac.uk/lensedquasars/index.html},
  note         = {Accessed: 2025-05-22}
}

@ARTICLE{2020_Pyhammer,
       author = {{Roulston}, Benjamin R. and {Green}, Paul J. and {Kesseli}, Aurora Y.},
        title = "{Classifying Single Stars and Spectroscopic Binaries Using Optical Stellar Templates}",
      journal = {\apjs},
         year = 2020,
        month = aug,
       volume = {249},
       number = {2},
          eid = {34},
        pages = {34},
          doi = {10.3847/1538-4365/aba1e7},
archivePrefix = {arXiv},
       eprint = {2006.01199},
 primaryClass = {astro-ph.SR},
       adsurl = {https://ui.adsabs.harvard.edu/abs/2020ApJS..249...34R}
}

@article{Kostas_2025, doi = {10.21105/joss.09264}, url = {https://doi.org/10.21105/joss.09264}, year = {2025}, publisher = {The Open Journal}, volume = {10}, number = {116}, pages = {9264}, author = {Valeckas, Kostas and Fynbo, Johan Peter Uldall and Krogager, Jens-Kristian and Heintz, Kasper Elm}, title = {PyLongslit: a simple manual Python pipeline for processing of astronomical long-slit spectra recorded with CCD detectors}, journal = {Journal of Open Source Software} }

@article{Heintz_2018,
   title={A quasar hiding behind two dusty absorbers: Quantifying the selection bias of metal-rich, damped Lyαabsorption systems},
   volume={615},
   ISSN={1432-0746},
   url={http://dx.doi.org/10.1051/0004-6361/201731964},
   DOI={10.1051/0004-6361/201731964},
   journal={Astronomy \&amp; Astrophysics},
   publisher={EDP Sciences},
   author={Heintz, K. E. and Fynbo, J. P. U. and Ledoux, C. and Jakobsson, P. and Møller, P. and Christensen, L. and Geier, S. and Krogager, J.-K. and Noterdaeme, P.},
   year={2018},
   month=jul, pages={A43} }

@INPROCEEDINGS{Ester_1996,
       author = {{Ester}, Martin and {Kriegel}, Hans-Peter and {Sander}, J{\"o}rg and {Xu}, Xiaowei},
        title = "{A Density-Based Algorithm for Discovering Clusters in Large Spatial Databases with Noise}",
    booktitle = {Second International Conference on Knowledge Discovery and Data Mining (KDD'96). Proceedings of a conference held August 2-4},
         year = 1996,
       editor = {{Pfitzner}, D.~W. and {Salmon}, J.~K.},
        month = jan,
        pages = {226-331},
       adsurl = {https://ui.adsabs.harvard.edu/abs/1996kddm.conf..226E}
}

@misc{Harris_2010,
      title={A New Catalog of Globular Clusters in the Milky Way}, 
      author={William E. Harris},
      year={2010},
      eprint={1012.3224},
      archivePrefix={arXiv},
      primaryClass={astro-ph.GA},
      url={https://arxiv.org/abs/1012.3224}, 
}

@ARTICLE{Gaia_2023,
       author = {{Gaia Collaboration} and {Vallenari}, A. and {Brown}, A.~G.~A. and {Prusti}, T. and {de Bruijne}, J.~H.~J. and {Arenou}, F. and {Babusiaux}, C. and {Biermann}, M. and {Creevey}, O.~L. and {Ducourant}, C. and {Evans}, D.~W. and {Eyer}, L. and {Guerra}, R. and {Hutton}, A. and {Jordi}, C. and {Klioner}, S.~A. and {Lammers}, U.~L. and {Lindegren}, L. and {Luri}, X. and {Mignard}, F. and {Panem}, C. and {Pourbaix}, D. and {Randich}, S. and {Sartoretti}, P. and {Soubiran}, C. and {Tanga}, P. and {Walton}, N.~A. and {Bailer-Jones}, C.~A.~L. and {Bastian}, U. and {Drimmel}, R. and {Jansen}, F. and {Katz}, D. and {Lattanzi}, M.~G. and {van Leeuwen}, F. and {Bakker}, J. and {Cacciari}, C. and {Casta{\~n}eda}, J. and {De Angeli}, F. and {Fabricius}, C. and {Fouesneau}, M. and {Fr{\'e}mat}, Y. and {Galluccio}, L. and {Guerrier}, A. and {Heiter}, U. and {Masana}, E. and {Messineo}, R. and {Mowlavi}, N. and {Nicolas}, C. and {Nienartowicz}, K. and {Pailler}, F. and {Panuzzo}, P. and {Riclet}, F. and {Roux}, W. and {Seabroke}, G.~M. and {Sordo}, R. and {Th{\'e}venin}, F. and {Gracia-Abril}, G. and {Portell}, J. and {Teyssier}, D. and {Altmann}, M. and {Andrae}, R. and {Audard}, M. and {Bellas-Velidis}, I. and {Benson}, K. and {Berthier}, J. and {Blomme}, R. and {Burgess}, P.~W. and {Busonero}, D. and {Busso}, G. and {C{\'a}novas}, H. and {Carry}, B. and {Cellino}, A. and {Cheek}, N. and {Clementini}, G. and {Damerdji}, Y. and {Davidson}, M. and {de Teodoro}, P. and {Nu{\~n}ez Campos}, M. and {Delchambre}, L. and {Dell'Oro}, A. and {Esquej}, P. and {Fern{\'a}ndez-Hern{\'a}ndez}, J. and {Fraile}, E. and {Garabato}, D. and {Garc{\'\i}a-Lario}, P. and {Gosset}, E. and {Haigron}, R. and {Halbwachs}, J.-L. and {Hambly}, N.~C. and {Harrison}, D.~L. and {Hern{\'a}ndez}, J. and {Hestroffer}, D. and {Hodgkin}, S.~T. and {Holl}, B. and {Jan{\ss}en}, K. and {Jevardat de Fombelle}, G. and {Jordan}, S. and {Krone-Martins}, A. and {Lanzafame}, A.~C. and {L{\"o}ffler}, W. and {Marchal}, O. and {Marrese}, P.~M. and {Moitinho}, A. and {Muinonen}, K. and {Osborne}, P. and {Pancino}, E. and {Pauwels}, T. and {Recio-Blanco}, A. and {Reyl{\'e}}, C. and {Riello}, M. and {Rimoldini}, L. and {Roegiers}, T. and {Rybizki}, J. and {Sarro}, L.~M. and {Siopis}, C. and {Smith}, M. and {Sozzetti}, A. and {Utrilla}, E. and {van Leeuwen}, M. and {Abbas}, U. and {{\'A}brah{\'a}m}, P. and {Abreu Aramburu}, A. and {Aerts}, C. and {Aguado}, J.~J. and {Ajaj}, M. and {Aldea-Montero}, F. and {Altavilla}, G. and {{\'A}lvarez}, M.~A. and {Alves}, J. and {Anders}, F. and {Anderson}, R.~I. and {Anglada Varela}, E. and {Antoja}, T. and {Baines}, D. and {Baker}, S.~G. and {Balaguer-N{\'u}{\~n}ez}, L. and {Balbinot}, E. and {Balog}, Z. and {Barache}, C. and {Barbato}, D. and {Barros}, M. and {Barstow}, M.~A. and {Bartolom{\'e}}, S. and {Bassilana}, J.-L. and {Bauchet}, N. and {Becciani}, U. and {Bellazzini}, M. and {Berihuete}, A. and {Bernet}, M. and {Bertone}, S. and {Bianchi}, L. and {Binnenfeld}, A. and {Blanco-Cuaresma}, S. and {Blazere}, A. and {Boch}, T. and {Bombrun}, A. and {Bossini}, D. and {Bouquillon}, S. and {Bragaglia}, A. and {Bramante}, L. and {Breedt}, E. and {Bressan}, A. and {Brouillet}, N. and {Brugaletta}, E. and {Bucciarelli}, B. and {Burlacu}, A. and {Butkevich}, A.~G. and {Buzzi}, R. and {Caffau}, E. and {Cancelliere}, R. and {Cantat-Gaudin}, T. and {Carballo}, R. and {Carlucci}, T. and {Carnerero}, M.~I. and {Carrasco}, J.~M. and {Casamiquela}, L. and {Castellani}, M. and {Castro-Ginard}, A. and {Chaoul}, L. and {Charlot}, P. and {Chemin}, L. and {Chiaramida}, V. and {Chiavassa}, A. and {Chornay}, N. and {Comoretto}, G. and {Contursi}, G. and {Cooper}, W.~J. and {Cornez}, T. and {Cowell}, S. and {Crifo}, F. and {Cropper}, M. and {Crosta}, M. and {Crowley}, C. and {Dafonte}, C. and {Dapergolas}, A. and {David}, M. and {David}, P. and {de Laverny}, P. and {De Luise}, F. and {De March}, R.},
        title = "{Gaia Data Release 3. Summary of the content and survey properties}",
      journal = {\aap},
         year = 2023,
        month = jun,
       volume = {674},
          eid = {A1},
        pages = {A1},
          doi = {10.1051/0004-6361/202243940},
archivePrefix = {arXiv},
       eprint = {2208.00211},
 primaryClass = {astro-ph.GA},
       adsurl = {https://ui.adsabs.harvard.edu/abs/2023A\&A...674A...1G}
}

@ARTICLE{Ducourant_2023,
       author = {{Ducourant}, C. and {Krone-Martins}, A. and {Galluccio}, L. and {Teixeira}, R. and {Le Campion}, J. -F. and {Slezak}, E. and {de Souza}, R. and {Gavras}, P. and {Mignard}, F. and {Guiraud}, J. and {Roux}, W. and {Managau}, S. and {Semeux}, D. and {Blazere}, A. and {Helmer}, A. and {Pourbaix}, D.},
        title = "{Gaia Data Release 3. Surface brightness profiles of galaxies and host galaxies of quasars}",
      journal = {\aap},
         year = 2023,
        month = jun,
       volume = {674},
          eid = {A11},
        pages = {A11},
          doi = {10.1051/0004-6361/202243798},
archivePrefix = {arXiv},
       eprint = {2206.14491},
 primaryClass = {astro-ph.GA},
       adsurl = {https://ui.adsabs.harvard.edu/abs/2023A\&A...674A..11D}
}

@article{Shu_2019,
    author = {Shu, Yiping and Koposov, Sergey E and Evans, N Wyn and Belokurov, Vasily and McMahon, Richard G and Auger, Matthew W and Lemon, Cameron A},
    title = {Catalogues of active galactic nuclei from Gaia and unWISE data},
    journal = {Monthly Notices of the Royal Astronomical Society},
    volume = {489},
    number = {4},
    pages = {4741-4759},
    year = {2019},
    month = {09},
    issn = {0035-8711},
    doi = {10.1093/mnras/stz2487},
    url = {https://doi.org/10.1093/mnras/stz2487},
    eprint = {https://academic.oup.com/mnras/article-pdf/489/4/4741/30042879/stz2487.pdf},
}

@ARTICLE{Assef_2018,
       author = {{Assef}, R.~J. and {Stern}, D. and {Noirot}, G. and {Jun}, H.~D. and {Cutri}, R.~M. and {Eisenhardt}, P.~R.~M.},
        title = "{The WISE AGN Catalog}",
      journal = {\apjs},
         year = 2018,
        month = feb,
       volume = {234},
       number = {2},
          eid = {23},
        pages = {23},
          doi = {10.3847/1538-4365/aaa00a},
archivePrefix = {arXiv},
       eprint = {1706.09901},
 primaryClass = {astro-ph.GA},
       adsurl = {https://ui.adsabs.harvard.edu/abs/2018ApJS..234...23A}
}

@misc{Flesch_2021,
      title={The Million Quasars (Milliquas) v7.2 Catalogue, now with VLASS associations. The inclusion of SDSS-DR16Q quasars is detailed}, 
      author={Eric Wim Flesch},
      year={2021},
      eprint={2105.12985},
      archivePrefix={arXiv},
      primaryClass={astro-ph.GA},
      url={https://arxiv.org/abs/2105.12985}, 
}

@ARTICLE{Gaia_2024,
    author = {{Gaia Collaboration} and {Krone-Martins}, A. and {Ducourant}, C. and {Galluccio}, L. and {Delchambre}, L. and {Oreshina-Slezak}, I. and {Teixeira}, R. and {Braine}, J. and {Le Campion}, J. -F. and {Mignard}, F. and {Roux}, W. and {Blazere}, A. and {Pegoraro}, L. and {Brown}, A.~G.~A. and {Vallenari}, A. and {Prusti}, T. and {de Bruijne}, J.~H.~J. and {Arenou}, F. and {Babusiaux}, C.},
    title = "{Gaia Focused Product Release: A catalogue of sources around quasars to search for strongly lensed quasars}",
    journal = {\aap},
    year = 2024,
    month = may,
    volume = {685},
    eid = {A130},
    pages = {A130},
    doi = {10.1051/0004-6361/202347273},
archivePrefix = {arXiv},
       eprint = {2310.06295},
 primaryClass = {astro-ph.GA},
       adsurl = {https://ui.adsabs.harvard.edu/abs/2024A\&A...685A.130G}
}

@article{Heintz_2020,
    title = "Spectroscopic classification of a complete sample of astrometrically-selected quasar candidates using Gaia DR2",
    author = "Heintz, {K. E.} and Fynbo, {J. P. U.} and Geier, {S. J.} and P. Moller and J-K Krogager and C. Konstantopoulou and {de Burgos}, A. and L. Christensen and Steinhardt, {C. L.} and B. Milvang-Jensen and P. Jakobsson and E. Hog and Arvedlund, {B. E. H. K.} and Christiansen, {C. R.} and Hansen, {T. B.} and Henriksen, {P. D.} and Kuszon, {K. B.} and McKenzie, {I. B.} and Mosekjaer, {K. A.} and Paulsen, {M. F. K.} and Sukstorf, {M. N.} and Wilson, {S. N.} and Orgaard, {S. K. K.}",
    year = "2020",
    month = nov,
    day = "24",
    doi = "10.1051/0004-6361/202039262",
    language = "English",
    volume = "644",
    journal = "Astronomy \& Astrophysics",
    issn = "0004-6361",
    publisher = "E D P Sciences",
}

%\section{Appendix}
%\input{6. Appendix}

\end{document}